\documentclass[
 reprint,
nofootinbib,
 amsmath,amssymb,
 aps,
floatfix 
]{revtex4-1}

\usepackage{graphicx}% Include figure files
\usepackage{dcolumn}% Align table columns on decimal point
\usepackage{bm}% bold math
\usepackage[colorinlistoftodos]{todonotes}
\usepackage[export]{adjustbox}%
\usepackage{natbib}

\renewcommand{\Pr}[1]{\ensuremath{\mathrm{Pr}\!\left\{#1\right\}}}

\begin{document}

\title{Contest models highlight inherent inefficiencies \\ of scientific funding competitions}

\author{Kevin Gross}
\email{krgross@ncsu.edu}
\affiliation{Department of Statistics \\ North Carolina State University \\ Raleigh, NC USA}

\author{Carl T. Bergstrom}
\email{cbergst@u.washington.edu}
\affiliation{Department of Biology \\ University of Washington \\ Seattle, WA USA} 

\date{\today}

\begin{abstract}
Scientific research funding is allocated largely through a system of soliciting and ranking competitive grant proposals.  In these competitions, the proposals themselves are not the deliverables that the funder seeks, but instead are used by the funder to screen for the most promising research ideas.  Consequently, some of the funding program's impact on science is squandered because applying researchers must spend time writing proposals instead of doing science.  To what extent does the community's aggregate investment in proposal preparation negate the scientific impact of the funding program? Are there alternative mechanisms for awarding funds that advance science more efficiently?  We use the economic theory of contests to analyze how efficiently grant proposal competitions advance science, and compare them with recently proposed, partially randomized alternatives such as lotteries.   We find that the effort researchers waste in writing proposals may be comparable to the total scientific value of the research that the funding supports, especially when only a few proposals can be funded.  Moreover, when professional pressures motivate investigators to seek funding for reasons that extend beyond the value of the proposed science (e.g., promotion, prestige), the entire program can actually hamper scientific progress when the number of awards is small.  We suggest that lost efficiency may be restored either by partial lotteries for funding, or by funding researchers based on past scientific success instead of proposals for future work.
\end{abstract}

\maketitle

\section*{Introduction}
Over the past fifty years, research funding in the United States has failed to keep pace with growth in scientific activity.  Funding rates in grant competitions have plummeted (Fig.~\ref{fig:nih}, \cite{cole1978,nih2015,deb2017,nsf2017}) and researchers spend far more time writing grant proposals than they did in the past \cite{geard2010}.  A large survey of top U.S. universities found that, on average, faculty devote 8\% of their total time --- and 19\% of their time available for research activities --- towards preparing grant proposals \cite{link2008}. Anecdotally, medical school faculty may spend fully half their time or more seeking grant funding \cite{siliciano2007,geard2010}.  While the act of writing a proposal may have some intrinsic scientific value \cite{barnett2013} --- perhaps by helping an investigator sharpen ideas --- much of the effort given to writing proposals is effort taken away from doing science \cite{alberts2014}. With respect to scientific progress, this time is wasted \cite{herbert2013}.  

Frustrated with the inefficiencies of the current funding system, some researchers have called for an overhaul of the prevailing funding model \cite{pagano2006,lawrence2009,graves2011,ioannidis2011,alberts2014,bollen2014,fang2016,avin2018b,vaesen2017}.  In particular, Fang \& Casadevall \cite{fang2016} recently suggested a partial lottery, in which proposals are rated as worthy of funding or not, and then a subset of the worthy proposals are randomly selected to receive funds.  Arguments in favor of a partial lottery include reduced demographic and systemic bias, increased transparency, and a hedge against the impossibility of forecasting how scientific projects will unfold \cite{fang2016}.  Indeed, at least three funding organizations --- New Zealand's Health Research Council and their Science for Technological Innovation program, as well as the Volkswagen Foundation \cite{avin2018a} --- have recently begun using partial lotteries to fund riskier, more exploratory science.

Compared with a proposal competition, a lottery permits more proposals to qualify for funding and thus lowers the bar that applicants must clear.  A lottery also offers a lower reward for success, as a successful proposal receives a chance at funding, not a guarantee of funding. Thus we expect that investigators applying to a partial lottery will invest less time and fewer resources in writing a proposal.  To a first approximation, then, a proposal competition funds high-value projects while wasting substantial researcher time on proposal preparation, whereas a partial lottery would fund lower-value projects on average but would reduce the time wasted writing proposals.  It is not obvious which system will have the greater net benefit for scientific progress.    

In this article, we study the merits and costs of traditional proposal competitions versus partial lotteries by situating both within the rich economic theory of {\em contests}.  In this theory, competing participants make costly investments (``bids'') in order to win one or more prizes \cite{moldovanu2001,vojnovic2016}.  Participants differ in key attributes, such as ability and opportunity cost, that determine their optimal strategies.  In an economics context, contests are often used by the organizer as a mechanism to elicit effort from the participants.  For example, TopCoder and Kaggle are popular contest platforms for tech firms (the organizers) to solicit programming or data-analysis effort from freelance workers (the participants).  However, because the participants' attributes influence their optimal strategies, the bids that participants submit reveal those attributes.  Thus, screening of participants often arises as a side effect.

In funding competitions, the organizer is a funding body and the participants are competing investigators. Investigators pitch project ideas of varying scientific value by preparing costly proposals.  However, unlike a traditional economic contest, the funding body's primary objective is to identify the most promising science, using proposals to screen for high-value ideas.  The funding body has little interest in eliciting work during the competition itself, as the proposals are not the deliverables the funder seeks.  All else equal, the funding agency would prefer to minimize the work that goes into preparing proposals, to leave as much time as possible for investigators to do science.  In this case, how should the funder organize the contest to support promising science without squandering much of the program's benefit on time wasted writing proposals?

Below, we pursue this question by presenting and analyzing a contest model for scientific funding competitions.  We first use the model to assess the efficiency of proposal competitions for promoting scientific progress, and ask how that efficiency depends on how many proposals are funded.  We next explore how efficiency is impacted when extra-scientific incentives such as professional advancement motivate scientists to pursue funding, and compare the efficiency of proposal competitions versus partial lotteries.  Finally, we reflect on alternative ways to improve the efficiency of funding competitions without adding intentional randomness to the award process.  All of our analyses focus on equilibrium behavior, and thus pertain most directly to long-standing funding competitions for which researchers can acquire experience that informs their future actions.

\section*{A contest model for scientific funding competitions}

Our model draws upon a framework for contests developed by Moldovanu \& Sela \cite{moldovanu2001}. In our application, a large number of scientists (or research teams) compete for grants to be awarded by a funding body.  The funder can fund a proportion $p$ of the competing investigators.  We call $p$ the payline, although $p$ could be smaller than the proportion of investigators who are funded if some investigators do not enter the competition.  

Project ideas vary in their scientific value, which we write as $v$, where $v \geq 0$.  In this case, scientific value combines the abilities of the investigator and the promise of the idea itself.  Although we do not assign specific units to $v$, scientific value can be thought of as some measure of scientific progress, such as the expected number of publications or discoveries.  We assume that the funder seeks to advance science by maximizing the scientific value of the projects that it funds, minus the value of the science that investigators forgo while writing proposals. However, the funder cannot observe the value of a project idea directly.  Instead the funder evaluates proposals for research projects, and awards grants to the top-ranked proposals.  Assume that proposals can be prepared to different strengths, denoted $x \geq 0$, with a larger value of $x$ corresponding to a stronger proposal.  A scientist with a project idea of value $v$ must decide how much effort to invest in writing a proposal, that is, to what strength $x$ her proposal should be prepared.  In our model, this decision is made by a cost-benefit optimization.  

On the benefit side, if a proposal is funded, the investigator receives a reward equal to the scientific value of the project, or $v$.  This reward is public, in the sense that it benefits both the investigator and the funder.  Receiving a grant may also bestow an extra-scientific reward on the recipient, such as prestige, promotion, or professional acclaim.  Write this extra-scientific reward as $v_0 \geq 0$.   This extra-scientific reward is private, as it benefits only the grant recipient, and not the funder.  Let $\eta(x)$ be the equilibrium probability that a proposal of strength $x$ is funded; $\eta(x)$ will be a non-decreasing function of $x$.  Thus, in expectation, an investigator with a project of value $v$ who prepares a proposal of strength $x$ receives a benefit of $(v_0 + v) \eta(x)$.

Preparing a grant proposal also entails a disutility cost, equal to the value of the science that the investigator could have produced with the time and resources invested in writing.  Let $c(v,x)$ give the disutility cost of preparing a proposal of strength $x$ for a project of value $v$.  Here, we study the case where $c(v,x)$ is a separable function of $v$ and $x$, so we set $c(v,x) = g(v) h(x)$.  Proposal competitions are effective screening devices because it is easier to write a strong proposal about a good idea than about a poor one.  Therefore, $g(v)$ is a decreasing function of $v$, i.e., $g'(v) < 0$.  For a given idea, it takes more work to write a stronger proposal, and thus $h'(x) > 0$.  Finally, we assume that preparing a zero-strength proposal is tantamount to opting out of the competition, which can be done at zero cost. Thus $h(0) = 0$. 

Preparing a proposal has some scientific value of its own through the sharpening of ideas that writing a proposal demands \cite{barnett2013}.  Let $k \in [0, 1)$ be the proportion of the disutility cost $c(v,x)$ that an investigator recoups by honing her ideas.  We call the recouped portion of the disutility cost the intrinsic scientific value of writing a proposal.  The portion of the disutility cost that cannot be recouped is scientific waste.

\begin{figure}[!ht]
	\begin{center}
		\includegraphics[width=0.6\linewidth]{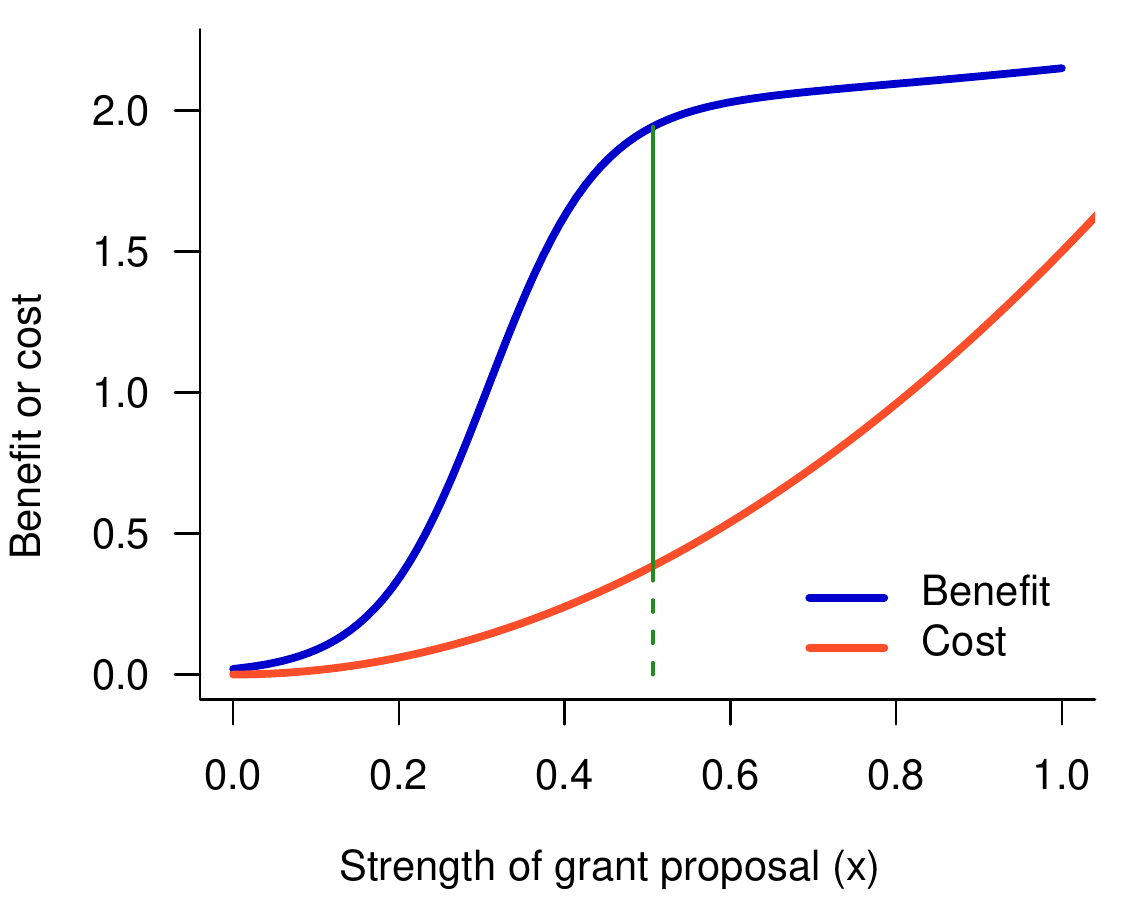}
		\caption{{\bf An investigator prepares her grant proposal to the strength that maximizes her payoff.} The blue curve shows the expected benefit to the investigator, which is determined by the project's value, any extra-scientific reward that the investigator receives from getting the grant, the probability of receiving funding, and the intrinsic value of writing a proposal.  The red curve shows the disutility cost of preparing a proposal.  The investigator's payoff is the difference between the benefit and the cost.  The vertical line shows the bid (eq.~\ref{eq:max-problem}) --- the proposal strength that maximizes the payoff.  At the bid, the ratio of the payoff (given by the length of the solid vertical line) to the cost (given by the length of the dashed vertical line) gives the investigator's return on her investment.}
		\label{fig:basic-setup}
	\end{center}
\end{figure}

All told, the total benefit to the investigator of preparing a proposal to strength $x$ is $(v_0 + v) \eta(x) + k c(v, x)$, and the total cost is $c(v, x)$.  The difference between the benefit and the cost is the investigator's payoff.  The investigator's optimal proposal (or, in economic terms, her ``bid'') maximizes this payoff (Fig.~\ref{fig:basic-setup}):
\begin{equation}
b(v) =  \underset{x}{\operatorname{arg\ max}} \left\{ (v_0 + v) \eta(x) - (1 - k) c(v,x) \right\}.
\label{eq:max-problem}
\end{equation}
For simplicity, we assume that variation among projects is captured entirely in the distribution of $v$, which we write as $F(v)$.  We assume that $v_0$ and $k$ have common values shared by all investigators.  In the appendix we show that our results extend to cases where $v_0$ or $k$ vary among investigators, as long as they are perfectly correlated with $v$.

The challenge in finding the payoff-maximizing bid $b(v)$ is that the equilibrium probability of funding, $\eta(x)$, must be determined endogeneously, in a way that is consistent with both the payline $p$ and the distribution of bids that investigators submit.  In Appendix S1, we follow Hoppe {\em et al.} \cite{hoppe2009} to show that, at equilibrium, the bid function is given by
\begin{equation}
b(v) = h^{-1} \left[\dfrac{1}{1-k}\int_0^v \frac{v_0 + t}{g(t)} \, \xi'(t) \, dt \right].
\label{eq:bid}
\end{equation}
In eq.~\ref{eq:bid}, $\xi(v) = \eta(b(v))$ is the equilibrium probability that an idea of value $v$ is funded.  The particular form of $\xi(v)$ depends on how much randomness is introduced during the review process, which we discuss below.  

By comparison, Moldovanu \& Sela \cite{moldovanu2001} considered a contest with a small number of competitors, in which the contest's judges observe $x$ directly.  In their set-up, each contestant is uncertain about the strength of her competition (that is, her competitors' types, $v$), but she can be certain that the strongest bid will win the top prize.  In our case, we assume that the applicant pool is large enough that the strength of the competition (i.e., the distribution of $v$ among the applicants) is predictable.  However, the funding agency does not observe $x$ directly, but instead convenes a review panel to assess each proposal's strength.  Variability among reviewers' opinions then introduces an element of chance into which proposals get funded.

\subsection*{Scientific efficiency}

We use the model to explore how efficiently the grant competition advances science. From the perspective of an individual investigator, the investigator's return on her investment (ROI) is the ratio of her payoff to the cost of her bid:
\begin{equation}
\mbox{Investigator's ROI} = \dfrac{(v_0 + v) \eta(b(v)) - (1 - k) c(v,b(v))}{c(v,b(v))}.
\label{eq:roi-agent}
\end{equation}
An investigator will never choose to write a proposal that generates a negative payoff, because she can always obtain a payoff of 0 by opting out.  (If the investigator opts out, eq.~\ref{eq:roi-agent} evaluates to 0/0, in which case we define her ROI to be 0.)  Thus, an investigator's equilibrium ROI must be $\geq 0$.  

To analyze the funding program's impact on scientific progress as a whole, we compare the total value of the science that the funding program supports with the total value of the science that has been squandered preparing proposals.  Of course, both of these quantities will be confounded with the number of grants that are funded, so we standardize to a per-funded-proposal basis.  In notation, the average scientific value per funded proposal is
\begin{equation}
\dfrac{1}{p} \int v \eta(b(v)) \, dF(v),
\label{eq:avg-value}
\end{equation}
and the average scientific waste per funded proposal is
\begin{equation}
\dfrac{1}{p} \int (1 - k) c(v,b(v)) \, dF(v).
\label{eq:avg-waste}
\end{equation}  
We will refer to the difference between these two quantities as the scientific gain (or loss, should it be negative) per funded proposal, which is our measure of the funding program's scientific efficiency. 

Note that while an investigator will never enter a grant competition against her own self interest, there is no guarantee that the scientific value per funded proposal will exceed the scientific waste.  This is because the investigator's payoff includes private, extra-scientific rewards obtained by winning a grant ($v_0$), and (in our accounting, at least) these extra-scientific rewards do not benefit the funding agency.  If extra-scientific motivations for winning grants are large enough, investigators may enter a grant competition even when doing so decreases their scientific productivity.  If enough investigators are motivated accordingly, then the scientific progress sacrificed to writing proposals could exceed the scientific value of the funding program.  In this case, the grant competition would operate at a loss to science, and the funding agency could do more for science by eschewing the proposal competition and spreading the money evenly among active researchers in the field, or by giving the money to researchers selected entirely at random.

\section*{Analysis and numerical results}

We illustrate the model's behavior by choosing a few possible sets of parameter values.  Our parameter choices are not directly informed by data.  Thus, while the numerical examples illustrate the model's possible behavior, we highlight the results that are guaranteed to hold in general.  Throughout, we use the following baseline set of parameters. We assume that the project values, $v$, have a triangular distribution ranging from $v_{\min} = 0.25$ to $v_{\max} = 1$ with a mode at $v_{\min}$, such that low-value ideas are common and high-value ideas are rare (i.e., $F(v) = 1 - (16/9)(1-v)^2$). For the cost function, we choose $c(v, x) = x^2 / v$.  We choose a convex dependence on $x$ to suggest that the marginal cost of improving a proposal increases as the proposal becomes stronger.  We assume that the intrinsic scientific value of writing a proposal allows investigators to recoup $k = 1/3$ of the disutility cost of proposal preparation.  We first explore the case when investigators are motivated purely by the scientific value of their projects ($v_0 = 0$), and then introduce extra-scientific benefits ($v_0 = 0.25$).  In the Supporting Information, we provide parallel results with two alternative parameter sets.

The evaluation process by which review panels rank proposals introduces a layer of randomness to the awarding of grants \cite{cole1981,fogelholm2012,pier2018}.  To capture noisy assessment, we use a bivariate \emph{copula} \cite{nelsen2006} to specify the joint distribution of a proposal's actual quantile, and its quantile as assessed by the funding agency's review panel.  A bivariate copula is a probability distribution on the unit square that has uniformly distributed marginals, as all quantiles must.  We use a Clayton copula \cite{clayton1978}, which allows for accurate assessment of weak proposals, but noisier assessment of strong proposals (Fig.~\ref{fig:copula}). This choice is motivated by the pervasive notion that review panels can readily distinguish strong proposals from weak ones, but struggle to discriminate among strong proposals \cite{scheiner2013,fang2016,pier2018}.  A Clayton copula has a single parameter ($\theta$) that controls how tightly its two components are correlated.  Rather arbitrarily, we use $\theta = 10$ in the baseline parameter set.  The Clayton copula has the important property that a proposal's probability of funding increases monotonically as its strength increases, regardless of the payline.  Thus, we exclude the possibility that panels systematically favor weaker proposals. By using a copula, we implicitly assume that $\eta(x)$ depends on $x$ only through its rank. In Appendix S1, we show how a copula leads to an equation for $\xi'(v)$, which can then be plugged in to eq.~\ref{eq:bid}.   

\begin{figure}[!ht]
	\begin{center}
		\includegraphics[width=0.9\linewidth]{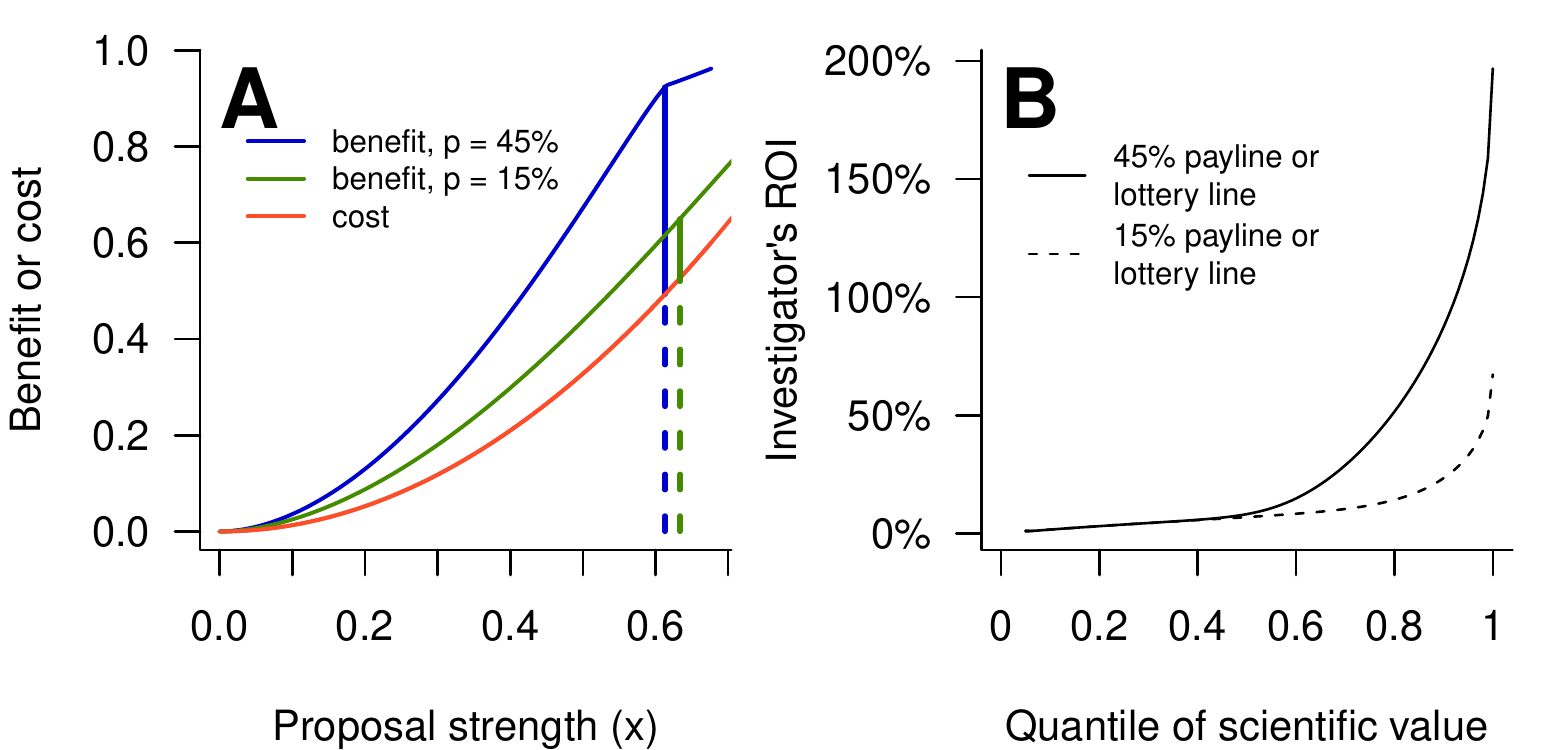}
		\caption{{\bf Diminishing paylines reduce investigators' returns on their investments in a proposal competition.}  A: Equilibrium benefit (blue or green) and cost (red) curves for an investigator with a project at the 90th percentile of $v$ in a proposal competition.  The blue and green curves show the benefit at a 45\% and 15\% payline, respectively.  Vertical lines show the investigator's equilibrium bid, with solid portions giving the investigator's payoff, and dashed portions showing the cost.  The corner in the benefit curves appears at the strongest proposal submitted.  B: An investigator's ROI (eq.~\ref{eq:roi-agent}) in a proposal competition with 45\% (solid line) or 15\% (dashed line) paylines, as a function of the quantile of the scientific value of her project, $F(v)$.  These curves also give the investigator's ROI in a partial lottery with a 45\% or 15\% lottery line, and any payline.  These results use the baseline parameters.}
		\label{fig:agent-a}
	\end{center}
\end{figure}

Fig.~\ref{fig:agent-a} shows numerical results for the baseline parameters at generous ($p = 45\%$) and low ($p = 15\%$) paylines.  In this particular case, investigators' payoffs fall faster than costs as paylines drop, leading to a reduced ROI for everyone at the lower payline (Fig.~\ref{fig:agent-a}B).  We will argue below that every investigator's ROI must inevitably fall when the payline becomes small (see Figs.~\ref{fig:agent-b}--\ref{fig:agent-c} for additional examples). 

\begin{figure}[!ht]
	\begin{center}
		\includegraphics[width=0.9\linewidth]{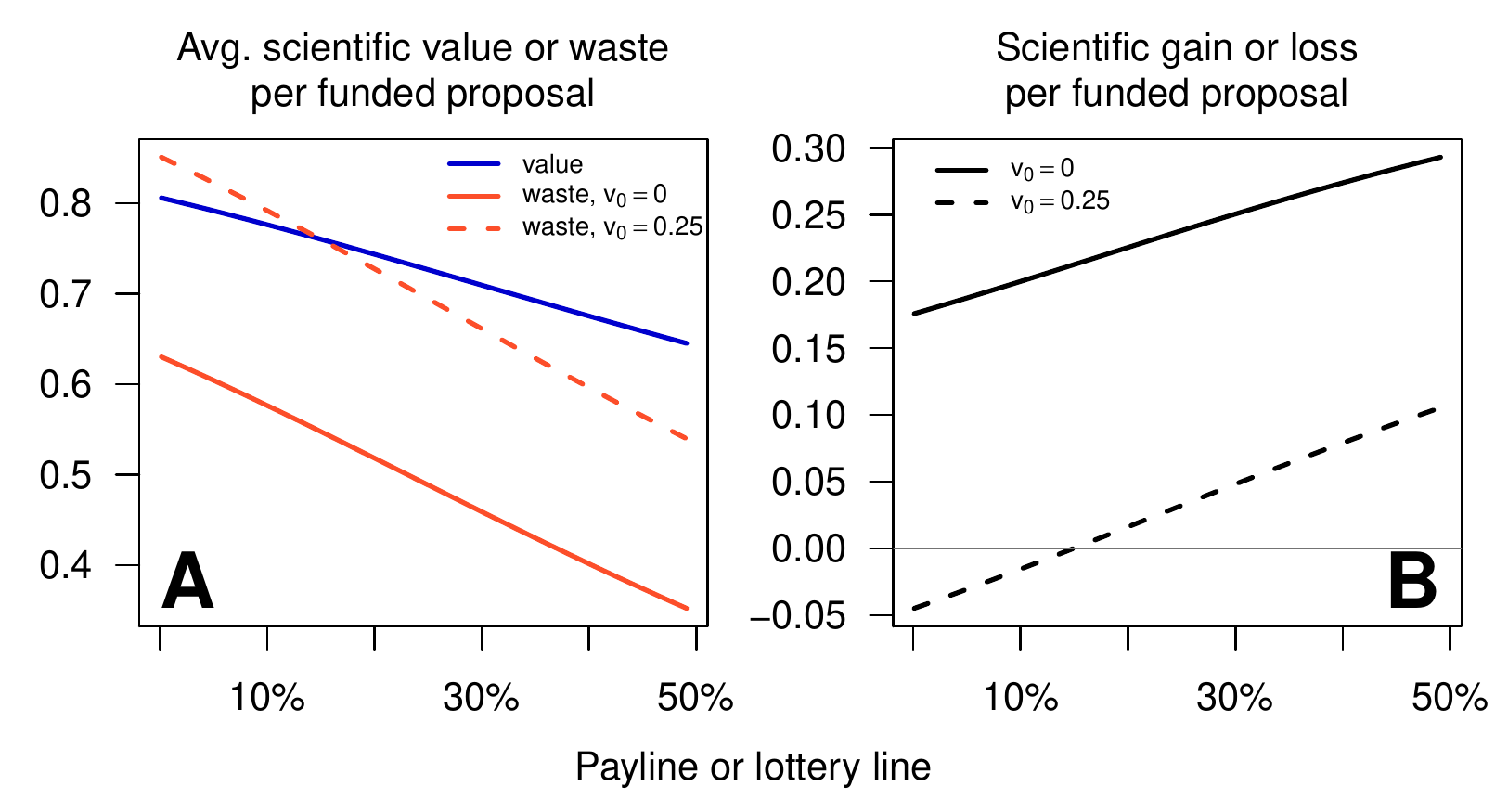}
		\caption{{\bf Both decreasing paylines and extra-scientific rewards to investigators reduce the scientific efficiency of the funding program.}  A: Both the average scientific value per funded proposal (blue line, eq.~\ref{eq:avg-value}) and the average waste per funded proposal (red lines, eq.~\ref{eq:avg-waste}) are higher for lower paylines, for the baseline parameter set and paylines $\leq 50\%$.  The solid red line shows costs when investigators are motivated only by the scientific value of the funded research ($v_0 = 0$); the dashed red line shows costs when investigators are additionally motivated by private, extra-scientific rewards ($v_0 = 0.25$).  Note that the vertical axis does not extend to 0.  B: The scientific gain or loss per funded proposal is lower for lower paylines, both in the absence ($v_0 = 0$, solid line) and presence ($v_0 = 0.25$, dashed line) of extra-scientific benefits to investigators.  Values of other parameters are as specified in the main text.  Identical results hold if the horizontal axis is re-interpreted as the proportion of proposals that qualify for a lottery, regardless of the payline.}
		\label{fig:principal-a}
	\end{center}
\end{figure}

From the funding agency's perspective, with our baseline parameters, both the average scientific value and average waste per funded proposal increase as the payline falls, for paylines below $50\%$ (Fig.~\ref{fig:principal-a}A). However, as the payline decreases, waste escalates more quickly than scientific value, reducing the scientific gain per funded project (Fig.~\ref{fig:principal-a}B).  This same result also appears in our alternative parameter sets (Fig.~\ref{fig:principal-b}--\ref{fig:principal-c}). We will argue below that the decline in scientific efficiency at low paylines is an inevitable if unfortunate characteristic of proposal competitions.
		
Clearly, quantitative details of the model's predictions depend on the parameter inputs.  To understand the robustness of these predictions, it helps to study the case where panels discriminate perfectly among proposals.  While perfect discrimination is obviously unrealistic in practice, it yields a powerful and general set of results that illuminate how the model behaves when discrimination is imperfect.  Numerical results for perfect discrimination under the baseline parameter set appear in Figs.~\ref{fig:agent-perfect}--\ref{fig:principal-perfect}.

At equilibrium under perfect assessment, every project above a threshold value $v^\star=F^{-1}(1-p)$ will receive funding, and no project idea below this threshold will be funded.  Investigators with projects of value $v > v^\star$ all prepare proposals to the identical strength $x^\star = h^{-1}\left[(v_0 + v^\star) / ( (1 - k) g(v^\star)  ) \right]$, and are funded with certainty.  Investigators with projects of value $v < v^\star$ opt out (Fig.~\ref{fig:agent-perfect}).  All of the subsequent results follow.  (Details appear in Appendix S1.)  First, as paylines drop, all investigators realize either a diminishing or zero ROI, because investigators who remain in the competition must pay a higher cost for a reduced payoff.  Second, the average scientific value per funded proposal must increase as paylines drop, because only the highest-value projects are funded under low paylines.  Third, in the limiting case when only one of many proposals can be funded (technically, the limit as $p$ approaches 0 from above), the scientific value and scientific waste associated with the last funded project converge, and science is no better off than if no grant had been given at all (Fig.~\ref{fig:principal-perfect}).  

With perfect assessment, there is no general relationship between the scientific efficiency of a proposal competition and the payline that holds across the full range of paylines (but see {Hoppe {\em et al.} \cite{hoppe2009} for a sharp result when the cost function is independent of $v$). Of course, we wouldn't expect scientific efficiency to decline monotonically with a falling payline, because there are likely to be some low-value projects that can be weeded out at low cost.  However, our last result above guarantees that the scientific gain per funded proposal must eventually vanish as the payline declines to a single award.

Returning to the reality of imperfect discrimination, as long as review panels do not systematically favor weaker proposals, noisy assessment changes little about these qualitative results.  That is, investigators' ROIs will drop as paylines fall, the average scientific value per funded proposal will increase as paylines decrease, and the scientific efficiency of the proposal competition must eventually decline as the payline approaches a single award.  But efficiency need not drop to zero. Perhaps counterintuitively, imperfect discrimination is a saving grace at low paylines. Noisy assessment discourages top investigators from pouring excessive effort into grant-writing as paylines fall, because the marginal benefit of writing an even better grant becomes small when review panels struggle to discriminate among top proposals.  Indeed, noisy assessment, unlike perfect discrimination,  allows a proposal competition to retain a positive impact on science even with a single funded grant (compare Figs.~\ref{fig:principal-a} and \ref{fig:principal-perfect}).  This result hints at the salutary nature of randomness at low paylines, which we will see more vividly when we consider lotteries below.

Thus  far, we have considered the case where investigators are motivated only by the scientific value of the projects proposed ($v_0 = 0$).  Now suppose that investigators are additionally motivated by the extra-scientific benefits of receiving a grant, such as professional advancement or prestige ($v_0 > 0$).  Eq.~\ref{eq:bid} shows that adding extra-scientific motivation will increase the effort that investigators devote to preparing grant proposals.  However, in our model at least, this extra effort has no bearing on which grants are funded, and thus does not affect the scientific value of the grants that are awarded.  Increasing scientific costs without increasing scientific value will clearly be detrimental to the funding program's scientific efficiency.  Extra-scientific benefits to investigators can even cause the entire funding program to operate at a loss to science when paylines are low (Fig.~\ref{fig:principal-a}).

\section*{Lotteries}

Our model can also be used to analyze the efficiency of a partial lottery for advancing science.  Suppose that a fraction $q \geq p$ of proposals qualify for the lottery, and each qualifying proposal is equally likely to be chosen for funding.  Call $q$ the ``lottery line''.  Now, the investigator's payoff is $(p / q) (v_0 + v) \eta_l(x) - (1 - k)c(v, x)$, where $\eta_l(x)$ is the equilibrium probability that the proposal qualifies for the lottery. In Appendix S1, we show that the investigator's bid is given by
\begin{equation}
b(v) = h^{-1} \left[ \dfrac{p}{q} \dfrac{1}{1 - k} \int_0^v \frac{v_0 + t}{g(t)} \, \xi'_l(t)   \, dt \right]
\label{eq:bid-lottery}
\end{equation}
where $\xi_l(v) = \eta_l(b(v))$.

Our major result for lotteries is that measures of scientific efficiency --- expressions \ref{eq:roi-agent}, \ref{eq:avg-value}, and \ref{eq:avg-waste} --- depend on the lottery line $q$ but are independent of the payline $p$ (proofs appear in Appendix S1).  This result follows from the fact that, in a lottery, each investigator's benefit and cost are proportional to $p$. Thus, an investigator's ROI and the scientific efficiency of the funding program are determined by the lottery line, but are not affected by the payline.  To illustrate, Fig.~\ref{fig:agent-comparison} compares an investigator's costs and benefits in a proposal competition with 45\%, 30\%, and 15\% paylines vs.\ a partial lottery with a $q = 45\%$ lottery line and the same three paylines.  The key feature of Fig.~\ref{fig:agent-comparison} is that the investigator's benefit curve in a partial lottery scales in such a way that her ROI is the same for any payline $\leq q$.  Consequently, a partial lottery with a lottery line of $q$ and any payline $\leq q$ achieves the same scientific efficiency as a proposal competition with a payline of $q$.  

Thus, our numerical results showing the investigator's ROI (Fig.~\ref{fig:agent-a}B) or the scientific efficiency (Fig.~\ref{fig:principal-a}) in a funding competition also show the efficiency of a lottery with the equivalent lottery line.  That is, a lottery in which 45\% of applicants qualify for the lottery has the same scientific efficiency as a proposal competition with a 45\% payline, regardless of the fraction of proposals chosen for the lottery that are ultimately funded. Thus, a lottery can restore the losses in efficiency that a proposal competition suffers as paylines become small.

\begin{figure}[!ht]
	\begin{center}
		\includegraphics[width=0.9\linewidth]{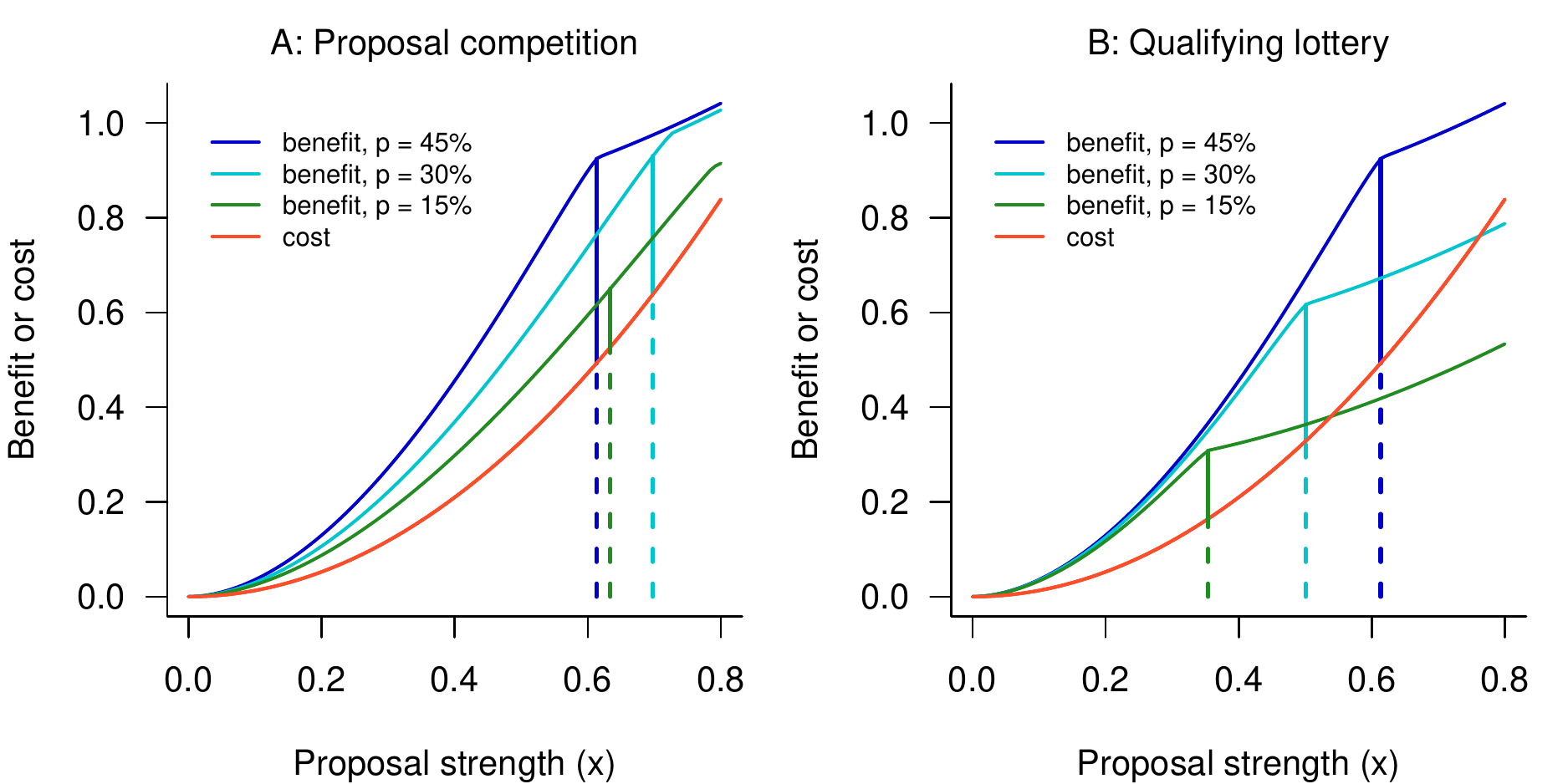}
		\caption{{\bf An investigator's ROI falls as the payline drops in a proposal competition, but is independent of the payline in a partial lottery.}  A: Benefit (blue or green) and cost (red) curves for an investigator with a project at the 90th percentile of $v$ in a proposal competition.  The dark blue, light blue, and green curves shows benefits with a 45\%, 30\%, and 15\% payline, respectively.  Vertical lines show the investigator's bid, with the length of the solid portion giving the payoff, and the length of the dashed portion giving the cost.  The investigator's ROI declines as the payline decreases.  (Note also that her effort does not vary monotonically with the payline.) B: The same investigator's benefit and cost curves in a partial lottery with a 45\% lottery line.  The investigator's ROI is the same for all paylines. These results use the baseline parameter set given in the main text.}
		\label{fig:agent-comparison}
	\end{center}
\end{figure}

In Appendix S1, we also analyze a more general type of lottery in which proposals are placed into one of a small number of tiers, with proposals in more selective tiers awarded a greater chance of funding \cite{brezis2007,graves2011,avin2018b}.  In a multi-tier lottery, the efficiency is entirely determined by the number of tiers and the relative probabilities of funding in each, and is independent of the payline. Numerical results (Fig.~\ref{fig:principal-lottery}) illustrate that the scientific value and waste of a multi-tier lottery fall in between those of a proposal competition and a single-tier lottery.  Thus, a multi-tier lottery offers an intermediate design that would partially reduce the waste associated with preparing proposals, while still allowing review panels to reward the best proposals with a higher probability of funding.  

\section*{Discussion}

Our major result is that proposal competitions are inevitably and inescapably inefficient mechanisms for funding science when the number of awards is smaller than the number of meritorious proposals.  The contest model presented here suggests that a partially randomized scheme for allocating funds --- that is, a lottery --- can restore the efficiency lost as paylines fall, albeit at the expense of reducing the average scientific value of the projects that are funded.  

Why does a lottery disengage efficiency from the payline, while a proposal competition does not?  For investigators, proposal competitions are, to a first approximation, all-or-nothing affairs --- an investigator only obtains a substantial payoff if her grant is funded.  At high paylines (or, more precisely, when the number of awards matches the number of high-value projects), investigators with high-value projects can write proposals that win funding at modest cost to themselves.  As the number of awards dwindles, however, competition stiffens.  Depending on the details of the assessment process, an investigator with a high-value project must either work harder for the same chance of funding, or work just as hard for a smaller chance of funding.  Either way, the return on her investment declines sharply.  Thus, a contest is most efficient at the payline that weeds out low-value projects, but does not attempt to discriminate among the high-value projects (e.g., Fig.~\ref{fig:principal-c}).  At lower paylines, however, the effort needed to signal which projects are most valuable begins to approach the value of those projects, making the funding program less worthwhile.

In a lottery, investigators do not compete for awards \emph{per se}, but instead compete for admission to the lottery.  The value to the investigator of being admitted to the lottery scales directly with the number of awards. It turns out that both the investigator's expected benefit and her costs of participation scale directly with the payline, and thus the payline has no effect on efficiency.  (In Appendix S1, we follow Hoppe {\em et al.} \cite{hoppe2009} to show that this scaling can be explained by the economic principle of revenue equivalence.)  If there are fewer awards than high-value projects, a lottery that weeds out the low-value projects but does not attempt to discriminate among high-value projects will facilitate scientific progress more efficiently than a contest.

% This analysis also shows that extra-scientific professional incentives to pursue grant funding can damage the scientific efficiency of a proposal competition.  To the extent that these extra-scientific incentives arise from administrators using grant success as a primary yardstick of professional achievement, perhaps one major benefit of adding explicit randomness to the funding mechanism would be to compel administrators to de-emphasize granting success in professional evaluations, and to look elsewhere for measures of scholarly success.  

Unfortunately, empirical comparisons between the efficiencies of funding competitions versus partial lotteries do not yet exist, to the best of our knowledge.  However, two recent anecdotes support our prediction that the waste in proposal competitions is driven by the strategic dynamics of the contest itself.  First, in 2012, the U.S. National Science Foundation's Divisions of Environmental Biology and Integrative Organismal Systems switched from a twice-annual, one-stage proposal competition to a once-annual, two-stage competition, in part to reduce applicants' workload.  However, the switch failed to reduce the applicants' aggregate workload meaningfully \cite{katz2017}, and the two-stage mechanism was subsequently abandoned.  Second, in 2014, the National Health and Medical Research Council of Australia streamlined the process of applying for their Project Grants, cutting the length of an application in half \cite{barnett2015}. However, researchers spent more time, not less, preparing proposals after the process had been streamlined, both individually and in aggregate \cite{barnett2015}.  Both of these experiences are consistent with our prediction that, in a proposal competition, the effort applicants expend is dictated by the value of funding to the applicants and the number of awards available, but does not depend on the particular format of the proposals.

A lottery is a radical alternative, and may be politically untenable \cite{barnett2016}.  If a lottery is not viable, an alternative approach to restoring efficiency is to design a contest where the effort given to competing for awards has more direct scientific value. For example, a contest that rewards good science in its completed form --- as opposed to rewarding well-crafted proposals that describe future science --- motivates the actual practice of good science, and will be less wasteful at low paylines \cite{ioannidis2011,alberts2014}.  Program officers could be given the discretion to allocate some funds by proactively scouting for promising researchers or projects. Of course, a contest based on completed science or scouting has its own drawbacks, including rich-getting-richer feedback loops, a risk of new barriers to entry for investigators from historically under-represented demographic groups, and the Goodhart's law phenomenon, whereby a metric that becomes a target ceases to be a good metric \cite{goodhart1984}.  Nevertheless, it is tantalizing to envision a world in which the resources that universities currently devote to helping researchers write proposals are instead devoted to helping researchers do science.

This analysis also shows that extra-scientific professional incentives to pursue grant funding can damage the scientific efficiency of a proposal competition.  As many of these extra-scientific incentives arise from administrators using grant success as a primary yardstick of professional achievement, perhaps one major benefit of adding explicit randomness to the funding mechanism would be to compel administrators to de-emphasize grant success in professional evaluations.  Alternatively, to the degree that administrators value and reward grant success because of the associated overhead funds that flow to the university, funding agencies could reduce waste by distributing overhead separately from funding awards.  Instead, perhaps overhead could be allocated based partially on the recent past productivity of investigators at qualifying institutions, among other possible criteria.  Disengaging overhead from individual grants would encourage administrators to value grants for the science those grants enable (as opposed to the overhead they bring), while allocating overhead based on institutions’ aggregate scientific productivity would motivate universities to help their investigators produce good science.

Funding agencies often have pragmatic reasons to emphasize the meritocratic nature of their award processes. However, our model also suggests that downplaying elements of a funding competition's structure that introduce randomness to funding decisions can increase scientific waste.  When applicants fail to recognize the degree to which the contest is already a lottery, they will over-invest effort in preparing proposals, to the detriment of science.  

This model does not account for all of the costs or scientific benefits of a proposal competition, including the costs of administering the competition, the time lost to reviewing grant proposals, or the benefit of building scientific community through convening a review panel.  Nonetheless, we suggest that the direct value of the science supported by funding awards and the disutility costs of preparing grant proposals are the predominant scientific benefits and costs of the usual proposal competition \cite{graves2011,katz2017}, and provide a useful starting point for a more detailed accounting. 

Our model also makes several simplifying assumptions, each of which may provide scope for interesting future work.  First, researchers pay a time cost to prepare a proposal, but receive money if the proposal is funded.  In our model, we have converted both time and money into scientific productivity, in order to place both on a common footing.  To be more explicit, though, scientific productivity requires both time and money (among other resources), and researchers may have vastly different needs for both.  In the Appendix, we show that our model can be formally extended to encompass researchers' different needs for time and money if the marginal rate of technical substitution (that is, the rate at which time and money can be exchanged without altering scientific productivity) is exactly correlated with the project's scientific value.  Our main results still hold in this case, as long as researchers with the best ideas do not value time so greatly that they write the weakest proposals.  A more general exploration of researchers' heterogeneous needs for time and money --- and of how researchers may adjust their portfolio of scientific activities when time or money is scarce --- provide ample opportunity for future work. 

%\change{This model also makes the simplifying assumption that time and money are substitutable resources with respect to how they enable scientific progress.}{Our model also makes several simplifying assumptions, each of which may provide scope for interesting future work.  First, this model assumes that time and money are substitutable resources with respect to how they enable scientific progress.  As we argue in the Appendix, the model does apply to scenarios where marginal rate of substitution for time vs. money differs among investigators, as long as the marginal rate of substitution is perfectly correlated with the project's scientific value.  (In this case, the marginal rate of substitution can be folded into the cost function.)}  \change{In reality, time and money can be complementary inputs into scientific production, with relative values that will vary hugely among investigators.  The present model could be extended by accounting for the sometimes complementary nature of time and money, and for researchers' different needs for both.}{In reality, however, time and money are often complements, not substitutes, and it would be interesting to extend the present model accordingly.}  

Second, our model assumes that the distribution of the scientific value ($v$) across possible projects is exogeneous to the structure of the funding competition.  This may not be the case if, for instance, a partial lottery encourages participation by investigators with unconventional views, reduces the psychological stigma of previous rejection \cite{fang2016}, or discourages investigators either who have succeeded under the traditional proposal-competition format or who perceive a lottery as riskier.  In reality, such feedback loops may endogenize the distribution of $v$.  Third, our model does not consider the savings that may accrue to investigators if they can submit a revised version of a rejected proposal to a different or subsequent competition. To a first approximation, submissions to multiple funders have the effect of increasing $p$, which can then be interpreted more generally as the proportion of ideas that get funded across all available funding programs.  Iterations of revision and resubmission to the same funding program are likely to have more complex effects on efficiency and waste.  Finally, our model is silent regarding whether many small or few large grants will promote scientific progress most efficiently, and is likewise silent about the factors that will influence this comparison.

%  Finally, we have assumed that other variables that motivate an investigator to seek grants (i.e., their extra-scientific reward for receiving a grant, the benefits they recoup from writing a grant, or their relative needs for time and money) are either constant, or vary among investigators in a way that is perfectly correlated project's scientific value.  While a full mathematical treatment of independent variation in these factors is beyond the scope of this manuscript, ...

To be sure, much more can be done to embellish this model.  However, the qualitative results --- that proposal competitions become increasingly inefficient as paylines drop, and that professional pressure on investigators to pursue funding exacerbates these inefficiencies --- are inherent to the structure of contests.  Partial lotteries and contests that reward past success present radical alternatives for allocating funds, and are sure to be controversial.  Nevertheless, whatever their other merits and drawbacks, these alternatives could  restore efficiency in distributing funds that has been lost as those funds have become increasingly scarce.

\section*{Acknowledgments}
We thank M. Lachmann for early discussions, A.\ Barnett and T.\ Bergstrom for useful feedback on an earlier draft, and B.\ Moldovanu for helping us articulate clearly the tension between proposal competitions and lotteries.  KG thanks the University of Washington Department of Biology for visitor support.

\bibliography{funding}

\newpage
\clearpage

\renewcommand\thefigure{S\arabic{figure}}  
\setcounter{figure}{0}    

\renewcommand\theequation{S\arabic{equation}}  
\setcounter{equation}{0}   

%\section*{Appendix S1: Mathematical Details}
\section*{Appendix S1: Mathematical Details}

\subsubsection*{Bid function}

The bid function $b(v)$ can be found following steps that are identical to the derivation of the first part of Proposition 8 in Hoppe {\em et al.} \cite{hoppe2009}.  We repeat that derivation here, almost fully verbatim.  

Begin by dividing the investigator's payoff function by by $(1 - k) g(v)$ to rescale the investigator's optimization problem (eq.~\ref{eq:max-problem}) :
\begin{equation}
b(v) =  \underset{x}{\operatorname{arg\ max}} \left\{ \frac{(v_0 + v)}{(1 - k) g(v)} \eta(x) - h(x) \right\}.
\end{equation}
Consider two projects of values $v_1$ and $v_2$, $v_1 > v_2$, with equilibrium bids $b(v_1)$ and $b(v_2)$, respectively.  The investigator with a project of value $v_1$ should not submit a bid as if her project had value $v_2$, and similarly the investigator whose project has value $v_2$ should not submit a bid as if her project had value $v_1$.  This yields:
\begin{eqnarray*}
\lefteqn{\frac{(v_0 + v_1)}{(1 - k) g(v_1)} \eta(b(v_1)) - h(b(v_1)) \geq } \hspace{0.5in} \\ & &  \frac{(v_0 + v_1)}{(1 - k) g(v_1)} \eta(b(v_2)) - h(b(v_2)) \\
\lefteqn{\frac{(v_0 + v_2)}{(1 - k) g(v_2)} \eta(b(v_2)) - h(b(v_2)) \geq } \hspace{0.5in} \\ & & \frac{(v_0 + v_2)}{(1 - k) g(v_2)} \eta(b(v_1)) - h(b(v_1)). 
\end{eqnarray*}
Rearrange each inequality to isolate $h(b(v_1)) - h(b(v_2))$ and divide through by $v_1 - v_2$ to give
\begin{widetext}
\begin{displaymath}
\dfrac{\frac{(v_0 + v_2)}{(1 - k) g(v_2)} (\eta(b(v_1)) - \eta(b(v_2)))}{v_1 - v_2} \leq \dfrac{h(b(v)) - h(b(\hat{v}))}{v_1 - v_2} \leq \dfrac{\frac{(v_0 + v_1)}{(1 - k) g(v_1)} (\eta(b(v_1)) - \eta(b(v_2)))}{v_1 - v_2}.
\end{displaymath}
\end{widetext}
Take the limit as $v_2 \rightarrow v_1$ to give 
\begin{equation}
\dfrac{d}{dv} h(b(v)) = \dfrac{v_0 + v}{(1 - k) g(v)} \dfrac{d}{dv} \eta(b(v)).
\label{eq:marginal}
\end{equation}
Essentially, after rescaling investigators' benefits and costs so that the cost function ($h(x)$) is the same for all investigators, eq.~\ref{eq:marginal} says that, at equilibrium, an investigator's marginal (re-scaled) cost and marginal (re-scaled) benefit of preparing an infinitesimally stronger proposal are equal \cite{hoppe2009}.  Proceeding with the derivation, multiply both sides of eq.~\ref{eq:marginal} by $dv$ to separate variables and integrate from 0 to $v$ to obtain
\begin{equation}
\int_0^v dh(b(t)) =  \int_0^v \frac{v_0 + t}{(1 - k) g(t)} \, \xi'(t) \, dt 
\end{equation}
where $\xi(v) = \eta(b(v))$.  Then use $h(b(0)) = h(0) = 0$ to give $\int_0^v dh(b(t)) = h(b(v))$, and thus 
\begin{equation}
h(b(v)) = \int_0^v \frac{v_0 + t}{(1 - k) g(t)} \, \xi'(t) \, dt .
\label{eq:bid-si}
\end{equation}
Take $h^{-1}$ on both sides to complete the derivation.

To make our derivation somewhat more general than in the main text, suppose that both $v_0$ and $k$ are themselves functions of $v$, such that the investigator's optimization problem now becomes
\begin{equation}
b(v) =  \underset{x}{\operatorname{arg\ max}} \left\{ \frac{(v_0(v) + v)}{(1 - k(v)) g(v)} \eta(x) - h(x) \right\}.
\end{equation}
We require the key condition that the quantity $\frac{(v_0(v) + v)}{(1 - k(v)) g(v)}$ is a strictly increasing function of $v$, so that investigators with higher value projects will submit stronger proposals.  This condition rules out the possibility that, say, variation in $v_0$ or $k$ is large enough that either replaces $v$ as the primary correlate of proposal strength.  When this key condition holds, the derivation above proceeds as before, leading to a bid function of
\begin{equation}
b(v) = h^{-1}\left[ \int_0^v \frac{v_0(t) + t}{(1 - k(t)) g(t)} \, \xi'(t) \, dt \right].
\end{equation}
In this case, variation in $v_0$ or $k$ among investigators may change the value of investigators' bids, but it does not change the rank of investigators' bids (that is, higher-value projects are still associated with stronger proposals).  Because we use a copula to capture noisy assessment of proposals, an investigator's probability of funding depends only on the rank of the investigator's bid, and thus the portfolio of funded projects does not change.

Finally, we can extend the model to accommodate researchers' differing needs for time and money, as long as the marginal rate of technical substitution for time vs.\ money is also a function of $v$.  In this case, we re-interpret the cost function $c(v, x) = g(v) h(x)$ as the time cost of writing a proposal, and we assume that the monetary cost of writing a proposal is negligible.  Let $\phi$ be a conversion factor that converts time into scientific productivity, such that the disutility cost of writing a proposal in terms of lost productivity is $\phi c(v, x)$.  Moreover, suppose $\phi$ is also a function of $v$.  Then, we can write the investigator's optimization problem as
\begin{equation}
b(v) =  \underset{x}{\operatorname{arg\ max}} \left\{ (v_0(v) + v) \eta(x) - (1 - k(v)) \phi(v) g(v) h(x) \right\}.
\label{eq:max-problem-si}
\end{equation}
As before, we require the key condition that $\frac{(v_0(v) + v)}{(1 - k(v)) \phi(v) g(v)}$ is a strictly increasing function of $v$, to ensure that proposal strength is positively correlated with scientific value at equilibrium.  Under this condition, the same steps give a bid function of
\begin{equation}
b(v) = h^{-1}\left[ \int_0^v \frac{v_0(t) + t}{(1 - k(t)) \phi(t) g(t)} \, \xi'(t) \, dt \right].
\end{equation}

To demonstrate that $b(v)$ maximizes the investigator's payoff (as opposed to minimizing it), we follow Moldovanu \& Sela's \cite{moldovanu2001} ``pseudo-concavity'' argument.  This argument requires that $b'(v) > 0$, which we establish first.  To do so, note that $h^{-1}$ in eq.~\ref{eq:bid} is an increasing function (because $h$ is an increasing function), and that $(v_0 + t) / g(t) > 0$ in the integrand of eq.~\ref{eq:bid}.  Thus, to show that $b'(v) > 0$, it suffices to show that $\xi'(v) > 0$, that is, that the probability of being funded increases as the value of the scientific project increases.  We expect this condition to hold under any reasonable model of how proposals are assessed.  Basic but tedious calculus establishes that it does hold for the Clayton copula that we describe below.

Having established that $b'(v) > 0$, the pseudo-concavity argument of Moldovanu \& Sela \cite{moldovanu2001} proceeds as follows.  Let $\varpi(v,x) = (v_0 + v) \eta(x) - (1 - k) g(v) h(x)$ be the payoff associated with a project of value $v$ and a proposal of quality $x$.  Let $\varpi_x = \partial \varpi(v,x) / \partial x$.  We claim that $\varpi_x > 0$ for $x < b(v)$, and $\varpi_x < 0$ for $x > b(v)$.  These claims, together with the continuity of $b(v)$, establish that $x = b(v)$ maximizes $\varpi(v, x)$. 

We first show that $\varpi_x > 0$ for $x < b(v)$.  Choose a value of $x < b(v)$, and let $v^\star$ be the value of a project that will generate a bid of $x$, that is, $b(v^\star) = x$.  Because $b'(v) > 0$, it follows that $v^\star < v$.  Simple differentiation gives
\begin{equation}
\varpi_x(v,x) = (v_0 + v) \eta'(x) - (1 - k) g(v) h'(x).
\end{equation}
Now differentiate $\varpi_x$ with respect to $v$ to give the mixed derivative
\begin{equation}
\varpi_{xv}(v,x) = \eta'(x) - (1 - k) g'(v) h'(x).
\end{equation}
Under the assumptions of our model, $\eta'(x) > 0$, $g'(v) < 0$, and $h'(x) > 0$; thus, $\varpi_{xv} > 0$.  Therefore, $\varpi_x$ is an increasing function of $v$, and thus $\varpi_x(v^\star, x) < \varpi_x(v, x)$.  By virtue of the fact that $b(v^\star) = x$, we have $\varpi_x(v^\star, x) = 0$.  Therefore, $\varpi_x(v, x) > 0$.

The proof that $\varpi_x < 0$ for $x > b(v)$ follows similarly.

\subsubsection*{Copulas for noisy assessment}  

A bivariate copula is simply a bivariate probability distribution on the unit square with uniform marginal distributions \cite{nelsen2006}.  Let $U = F(v)$ be the actual quantile of a proposal, and let $W$ be the assessed quantile.  The joint distribution of $U$ and $W$ is given by the copula $\mathcal{C}(u,w) = \Pr{U \leq u, W \leq w}$.  Given a value of $U$, the conditional distribution of $W$ given $U$ is $\mathcal{C}_{W|U}(u,w) = \Pr{W \leq w | U = u} = \partial \mathcal{C}(u,w) / \partial u$.  (Here we use the fact that $U$ is uniformly distributed on the unit interval.)  To find $\xi(v)$, evaluate $\mathcal{C}_{W|U}$ at $u = F(v)$ and $w = 1-p$ to find $1 - \xi(v)$, the probability that an idea of value $v$ is not funded.  Take the complement to find $\xi(v)$.  Differentiate with respect to $v$ to find $\xi'(v)$, which can then be plugged in to eq.~\ref{eq:bid}.

The distribution function for a Clayton copula \cite{clayton1978} is \cite[\S 4.2]{nelsen2006} 
\begin{equation}
\mathcal{C}(u,w) = \left(u^{-\theta} + w^{-\theta} - 1 \right)^{-1/\theta}.
\end{equation}
The parameter $\theta \geq 0$ controls the strength of the association between $U$ and $W$, with larger values of $\theta$ giving stronger associations (i.e., more accurate assessment of grant proposals). \\

\subsubsection*{Alternative parameter sets}  

To complement the example in the main text, we show numerical results for two alternative parameter sets.  In the first alternative set, scientific value is uniformly distributed across projects, the disutility cost increases linearly with proposal quality, and assessment is less precise than we assume in the baseline parameter set.  In this set, $v$ is uniformly distributed between 1/3 and 1.  We use $c(v,x) = x e^{-v}$ for the cost function, and we use $\theta = 5$ in the Clayton copula (Fig.~\ref{fig:copula}B).  

The second alternative parameter set captures a scenario where the pool of possible project values is bimodal, with many minimal-value projects and equally many maximal-value projects. In this alternative set, $v$ ranges from 1/2 to 1.  To construct the distribution of $v$, let $Y$ be a beta random variable with both shape parameters equal to 1/2.  Thus, $Y$ has a symmetric, U-shaped distribution on the unit interval. Then $v$ is given by $(1 + Y)/2$.  In this parameter set, we choose $c(v,x) = (1.5 - v) x^2$, and we set $\theta = 7.5$ in the Clayton copula.

Both alternative parameter sets use $k = 1/3$.

\subsubsection*{Perfect discrimination}

In the perfect discrimination case, we require that there is a maximum possible value of $v$, which we write $v_{\max}$.  The previous derivation of the bid function does not work for perfect discrimination, because $b(v)$ becomes a step function and thus is not differentiable.  Instead, let $v^\star$ denote the threshold value, that is, $v^\star = F^{-1}(1-p)$.  Under perfect discrimination, the threshold investigator will break even regardless of her bid.  Thus, 
\begin{equation}
\eta(x) = \begin{cases}
 \dfrac{g_k(v^\star)h(x)}{v_0 + v^\star} & x \leq x^\star \\
 1 & x > x^\star
 \end{cases}
\end{equation}
where 
\begin{equation}
x^\star = h^{-1}\left[\dfrac{v_0 + v^\star}{g_k(v^\star)}\right].
\end{equation} 
Consequently, it is straightforward to show that the bid function is 
\begin{equation}
b(v) = \begin{cases}
 x^\star  & v > v^\star \\
 0 & v < v^\star.
 \end{cases}
\end{equation}
The following results are all immediate.  First, as the payline drops, $v^\star$ increases, and hence $x^\star$ increases.  (Recall that $g$ is a strictly decreasing function, and $h^{-1}$ is a strictly increasing function.)  Thus, investigators with $v > v^\star$ experience a reduced payoff and increased costs, leading to a reduced ROI.  Second, the average scientific benefit per funded proposal, which can be written as
\begin{equation}
\int_{v^\star}^{v_{\max}} v \, dF(v) \bigg/ \int_{v^\star}^{v_{\max}} \, dF(v),
\end{equation}
increases as $v^\star$ increases.  Third, as $p$ approaches 0 from above, $v^\star$ approaches $v_{\max}$ from below.  Thus, in the limit, the bid function approaches
\begin{equation}
\lim_{p \rightarrow 0^+} b(v) = 
\begin{cases}
 h^{-1}\left[\dfrac{v_0 + v_{\max}}{g_k(v_{\max})}\right]  & v = v_{\max} \\
 0 & v < v_{\max}.
 \end{cases}
\end{equation}
Thus, the payoff to all investigators approaches 0 as $p$ approaches 0 from above.

\subsubsection*{Lotteries}

We consider the more general case of a multi-tier lottery.  Proposals deemed worthy of funding are placed into one of $z$ tiers, with tier 1 representing the highest-ranked proposals, etc.  Write the proportion of proposals in tier $i$ as $q_i$, and let $\pi_i$ represent the probability that a proposal placed in tier $i$ is funded, where $1 \geq \pi_1 > \pi_2 > \ldots > \pi_z > 0$.  We assume that the funding agency determines $q_1, q_2, \ldots, q_z$ and $\pi_1, \pi_2, \ldots, \pi_z$ in advance.  Because the payline is still $p$, we must have $\sum_{i=1}^z q_i \pi_i = p$.  The single-tier lottery proposed by Fang \& Casadevall \cite{fang2016} and others is a special case with $z=1$, with a probability of funding $\pi_1 = p / q_1$ in that tier. 

In a tiered lottery, the investigator's maximization problem becomes
\begin{equation}
b(v) =  \underset{x}{\operatorname{arg\ max}} \left\{ (v_0 + v) \sum_{i=1}^z \pi_i \eta_i(x) - (1 - k) c(v,x) \right\}
\label{eq:max-problem-lottery-si}
\end{equation}
where $\eta_i(x)$ is the probability that a proposal of quality $x$ is placed in tier $i$.  A similar derivation to the steps in eq.~\ref{eq:marginal}--\ref{eq:bid-si} yields the bid function
\begin{equation}
b(v) = h^{-1} \left[ \sum_{i=1}^z \dfrac{\pi_i}{1-k} \int_0^v \frac{v_0 + t}{g(t)} \, \xi'_i(t)   \, dt \right].
\label{eq:bid-lottery-si}
\end{equation}
where $\xi_i(v) = \eta_i(b(v))$.

We now show that the efficiency of a lottery depends entirely on the structure of the lottery, and is independent of the payline.  For multi-tier lotteries, we require that the ratios of the $\pi_i$'s --- the probabilities of funding in each tier --- are fixed.  To establish these ratios, write $\kappa_i = \pi_i / \pi_1$. The condition $\sum_{i=1}^z q_i \pi_i = p$ implies that the probability that a proposal in tier $i$ is funded is
\begin{equation}
\pi_i = \dfrac{p \kappa_i}{\sum_i \kappa_i q_i},
\end{equation}
as long as $p \leq \sum_i \kappa_i q_i$.  (If $p > \sum_i \kappa_i q_i$, then we would have $\pi_1 > 1$.)

All of our results follow from showing that an investigator's benefit and cost are proportional to $p$, and thus the payline $p$ cancels out of the efficiency calculations in eqq.~\ref{eq:roi-agent}--\ref{eq:avg-waste}.  The investigator's benefit from entering the competition is $(v_0 + v) \sum_{i=1}^z \pi_i \eta_i(x)$.  A simple substitution shows that this benefit is proportional to $p$:
\begin{eqnarray*}
(v_0 + v) \sum_{i=1}^z \pi_i \eta_i(x) & = & (v_0 + v) \sum_{i=1}^z \dfrac{p \kappa_i \eta_i(x) }{\sum_j \kappa_j q_j} \\
& = & p (v_0 + v) \dfrac{\sum_i \eta_i(x) \kappa_i}{\sum_j \kappa_j q_j}.
\end{eqnarray*}

To show that the investigator's cost is proportional to $p$, we have
\begin{eqnarray*}
c(v, b(v)) & = & g(v) h(b(v)) \\
& = & g(v) \sum_{i=1}^z \dfrac{\pi_i}{1 - k} \int_0^v \frac{v_0 + t}{g(t)} \, \xi'_i(t)   \, dt \\
& = & p \, g(v) \dfrac{\sum_i \kappa_i \int_0^v \frac{v_0 + t}{g(t)} \, \xi'_i(t)   \, dt}{(1 - k) \sum_j \kappa_j q_j} .
\end{eqnarray*}

It thus follows that the investigator's ROI (eq.~\ref{eq:roi-agent}), the average value per funded grant (eq.~\ref{eq:avg-value}), and the average waste per funded grant (eq.~\ref{eq:avg-waste}) are all independent of $p$.

Hoppe {\em et al.} \cite{hoppe2009} provide an argument based on the economic principle of revenue equivalence that explains why costs are independent of the payline in a lottery.  This argument applies both to proposal competitions and to lotteries of any structure, and it applies regardless of whether panels discriminate perfectly among proposals, or not.  The argument is most easily explained in a single-tier lottery with perfect discrimination, so we consider that setting.  First, for revenue equivalence to apply, we need to re-scale the model so that only benefits vary among investigators.  That is, re-scale the investigator's equilibrium benefit function to 
\begin{equation}
\dfrac{p}{q}\dfrac{(v_0 + v)}{(1-k)g(v)} \xi_l(v)
\label{eq:rescaled-benefit}
\end{equation}
and write her cost function as $h(x)$.  Having re-scaled the investigator's benefits and costs, the principle of revenue equivalence suggests that the (re-scaled) cost paid by an investigator will be exactly equal to the negative externality that her entrance into the competition creates, that is, the amount by which her entrance decreases the aggregate benefit of the competing investigators \cite{hoppe2009}.  With perfect discrimination, the threshold investigator (the last one to qualify for the lottery) has project value $v^\star = F^{-1}(1-q)$; every investigator with $v > v^\star$ qualifies for the lottery, while every investigator with $v < v^\star$ opts out.  Consider an investigator with an idea of value $v$.  If $v<v^\star$, then this investigator's entrance into the competition has no effect on other investigators' benefit, and thus the cost she pays is 0 (i.e., she opts out).  If the investigator has an idea of value $v > v^\star$, then she knocks one threshold investigator out of the lottery.  No other investigator's benefit changes.  Thus, the new investigator's presence decreases the aggregate benefit of the other investigators by 
\begin{equation}
\dfrac{p}{q}\dfrac{(v_0 + v^\star)}{(1-k)g(v^\star)}.
\end{equation}
This negative externality is exactly the re-scaled cost that she pays, i.e.,
\begin{equation}
h(b(v)) = \dfrac{p}{q}\dfrac{(v_0 + v^\star)}{(1-k)g(v^\star)}.
\end{equation}
Multiplying by $g(v)$ undoes the re-scaling to give the actual cost paid:
\begin{equation}
c(v, b(v)) = g(v) h(b(v)) = g(v) \dfrac{p}{q}\dfrac{(v_0 + v^\star)}{(1-k)g(v^\star)}.
\end{equation}

Two observations explain why the cost paid by the investigator is directly proportional to $p$.  First, the identity of the threshold investigator --- the one who is knocked out of the lottery when an investigator with a higher-value project enters --- is determined by $q$, not $p$.  (In a proposal competition, the threshold investigator is determined by $p$.)  Second, the threshold investigator's benefit --- and hence the negative externality imposed by the newly arriving investigator --- is directly proportional to $p$.  Thus, the (re-scaled) cost paid by any investigator will also be directly proportional to $p$.  Multiplying the cost by $g(v)$ to undo the rescaling does not change the direct proportionality to $p$.

With noisy assessment, and/or in a multi-tiered lottery, the negative externality that an investigator imposes on the field, and hence the (re-scaled) cost that she pays at equilibrium, integrates the amount by which her entry decreases the benefit of every investigator with a project value less than hers.  (This is one way to understand the bid functions in eq.~\ref{eq:bid}, \ref{eq:bid-lottery}, and \ref{eq:bid-lottery-si}.)  In a lottery, everyone's benefit is proportional to $p$, and hence each investigator's negative externality, and the cost that she pays, is proportional to $p$ as well.

\clearpage
% \section*{Supporting Information: Additional Figures}
\section*{Supporting Information Figures}

\begin{figure}[th!]
	\centering
		\includegraphics[width=0.8\linewidth]{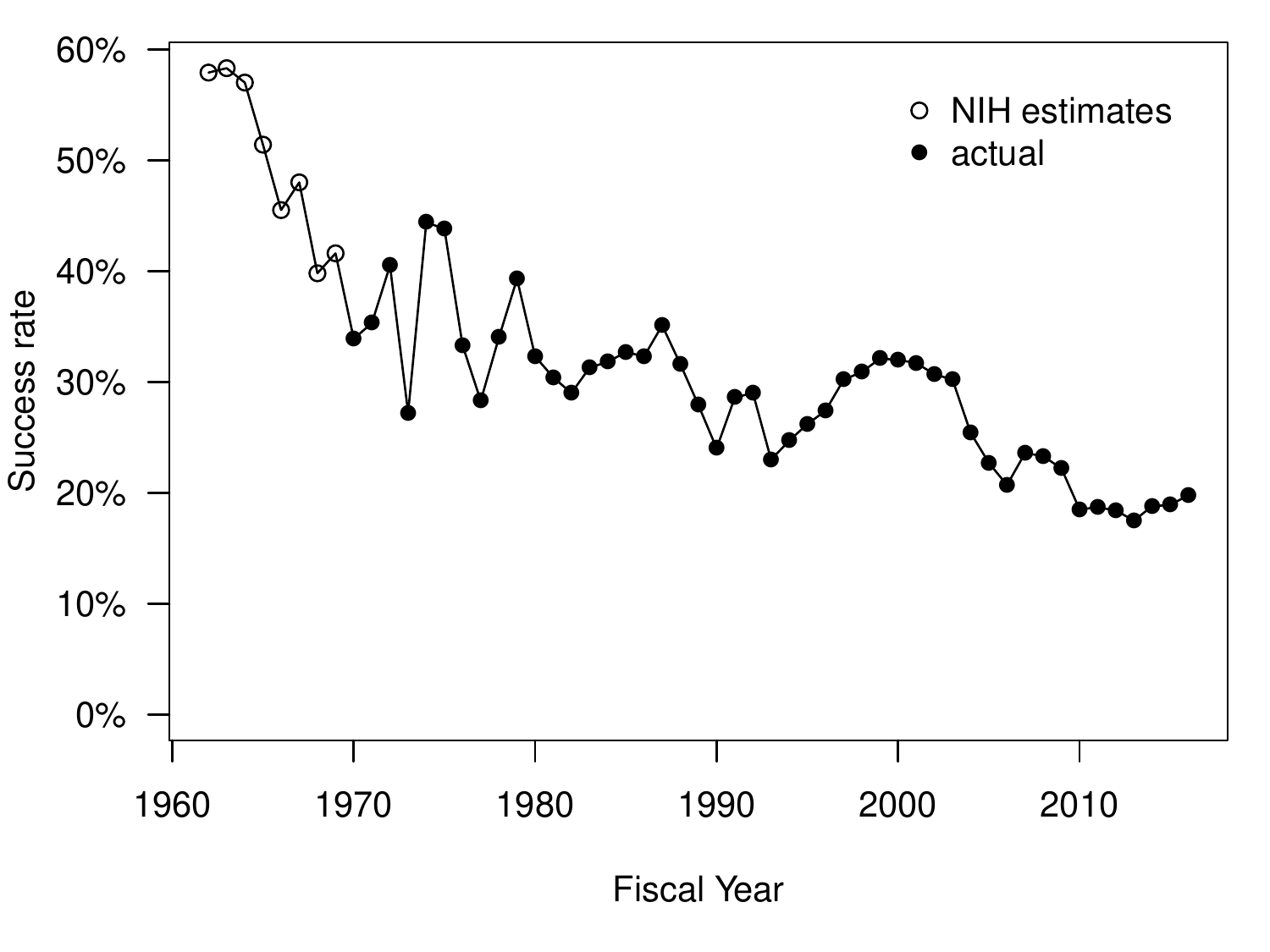}
		\caption{Funding rates of NIH R01 and equivalents from FY 1962 -- FY 2016.  Data from FY 1962 -- 2008 include R01, R23, R29, and R37 proposals, as reported by NIH's Office of Extramural Research \cite{nih}.  1962 -- 1969 data are NIH estimates.  Data for FY 2009 -- 2016 include R01 and R37 proposals, as reported by \cite{nih2015}.  (R01 and R37 provide the vast majority of proposals for earlier years.)  Data include new applications, supplements, and renewals, and the success rate is calculated as the number of proposals funded divided by the number of proposals reviewed.}
		\label{fig:nih}
\end{figure}

\begin{figure}
	\begin{center}
		\includegraphics[width=0.8\linewidth]{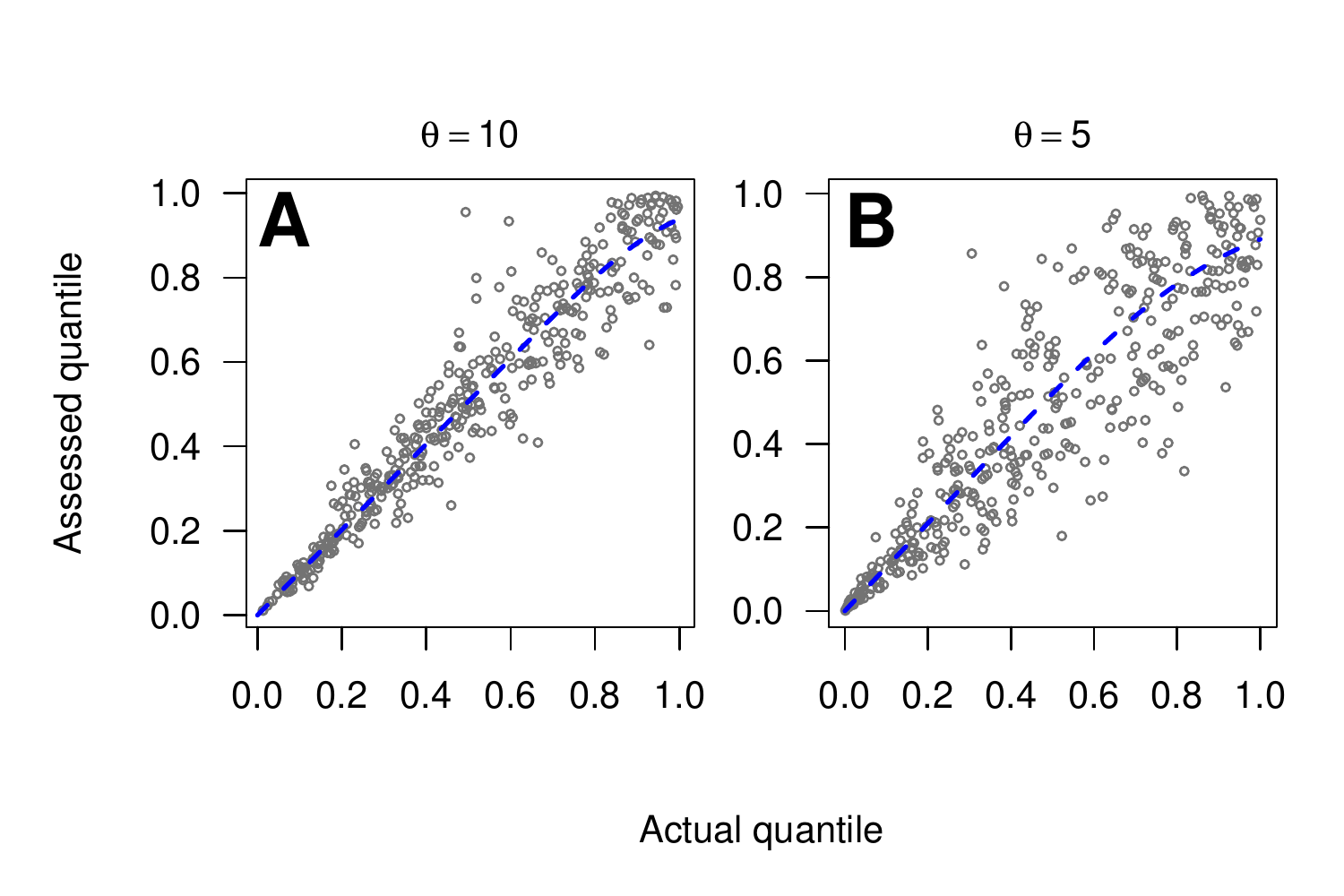}
		\caption{Random samples from copula distributions used to model error in assessment of grant proposals.  A: Clayton copula with $\theta = 10$. B: Clayton copula with $\theta = 5$. Blue dashed lines give the median of the assessed quantile as a function of the actual quantile.}
		\label{fig:copula}
	\end{center}
\end{figure}

\begin{figure}
	\begin{center}
		\includegraphics[width=0.8\linewidth]{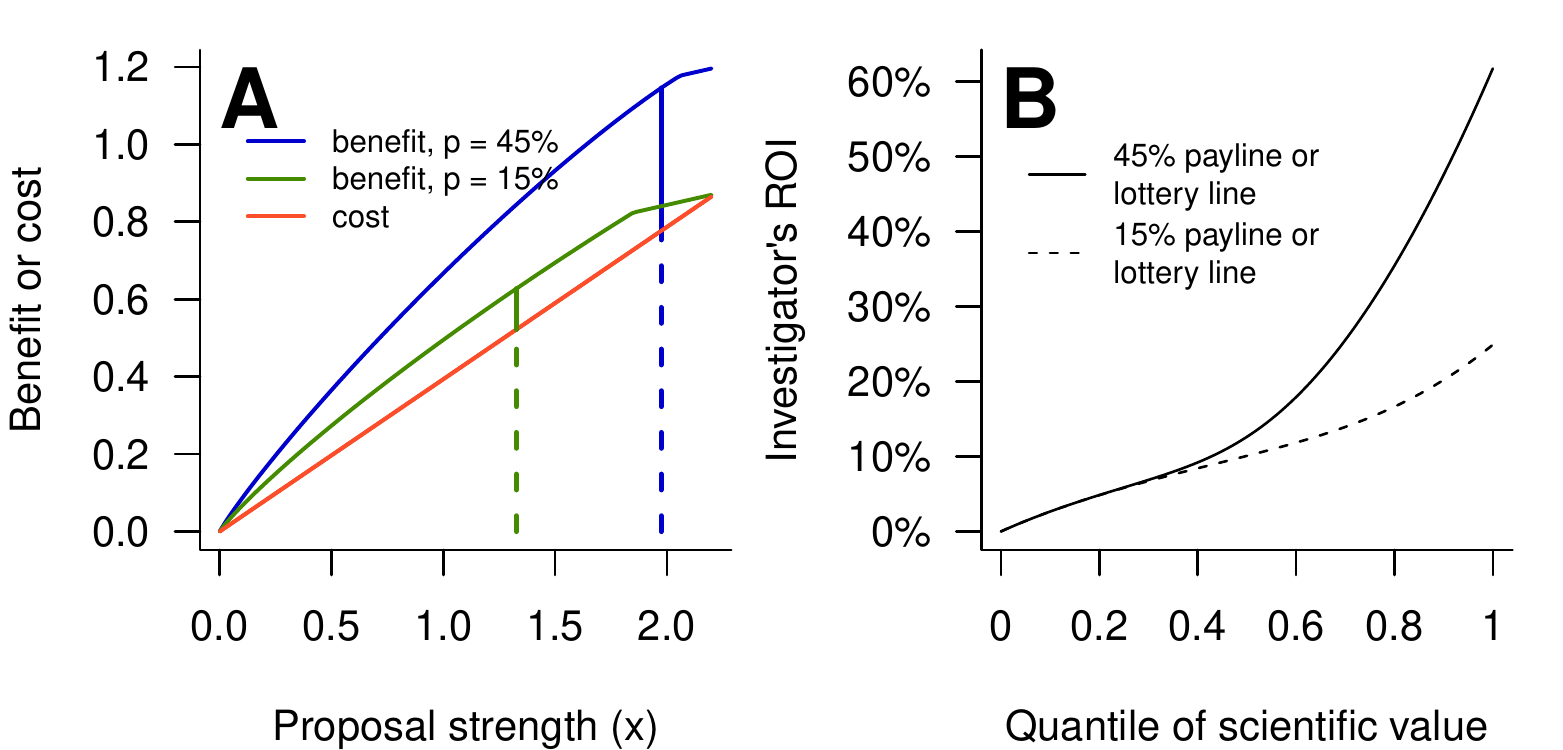}
		\caption{Parallel results to Fig.~\ref{fig:agent-a}, except with the first alternative parameter set given in the SI.}
		\label{fig:agent-b}
	\end{center}
\end{figure}

\begin{figure}
	\begin{center}
		\includegraphics[width=0.8\linewidth]{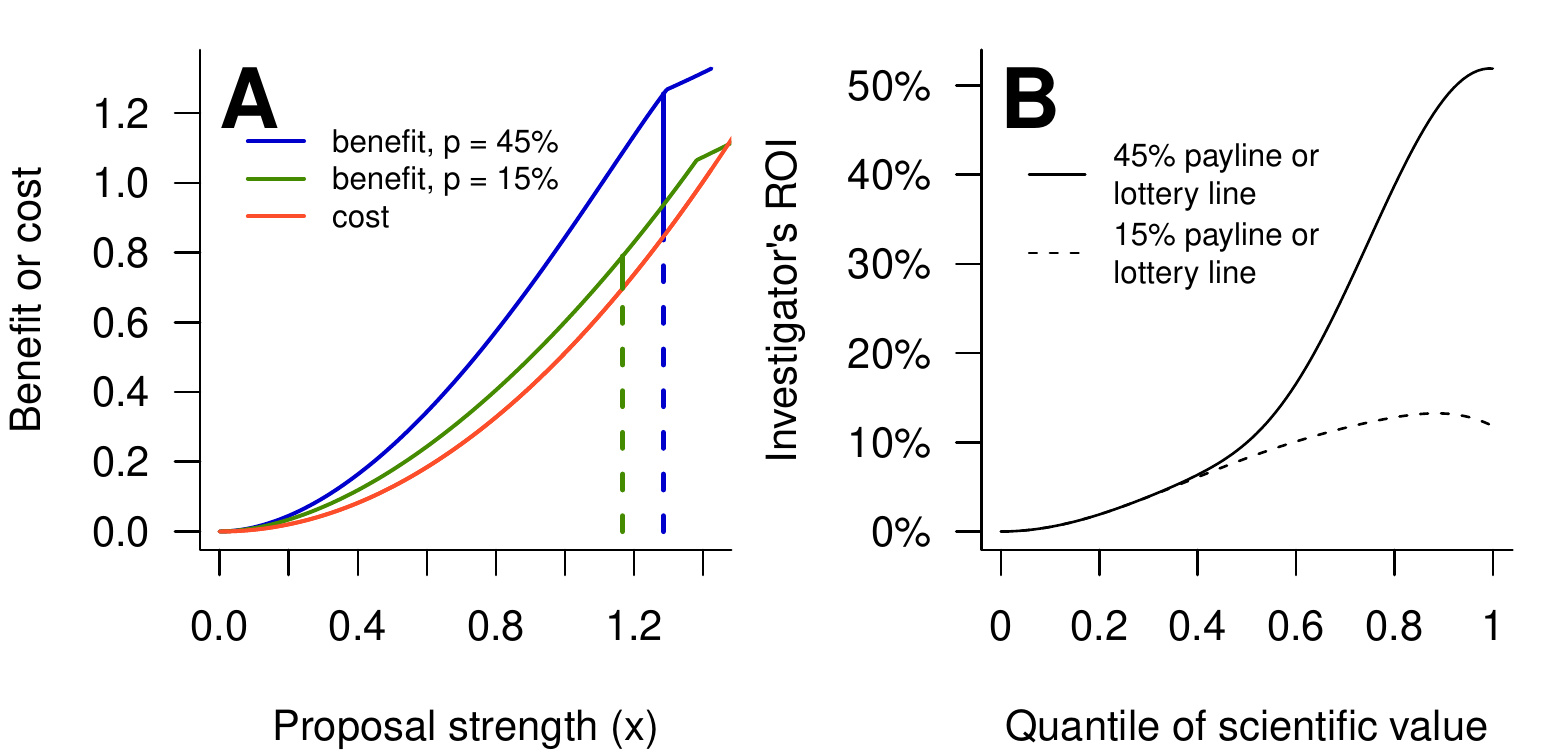}
		\caption{Parallel results to Fig.~\ref{fig:agent-a}, except with the second alternative parameter set.}
		\label{fig:agent-c}
	\end{center}
\end{figure}

\begin{figure}
	\begin{center}
		\includegraphics[width=0.8\linewidth]{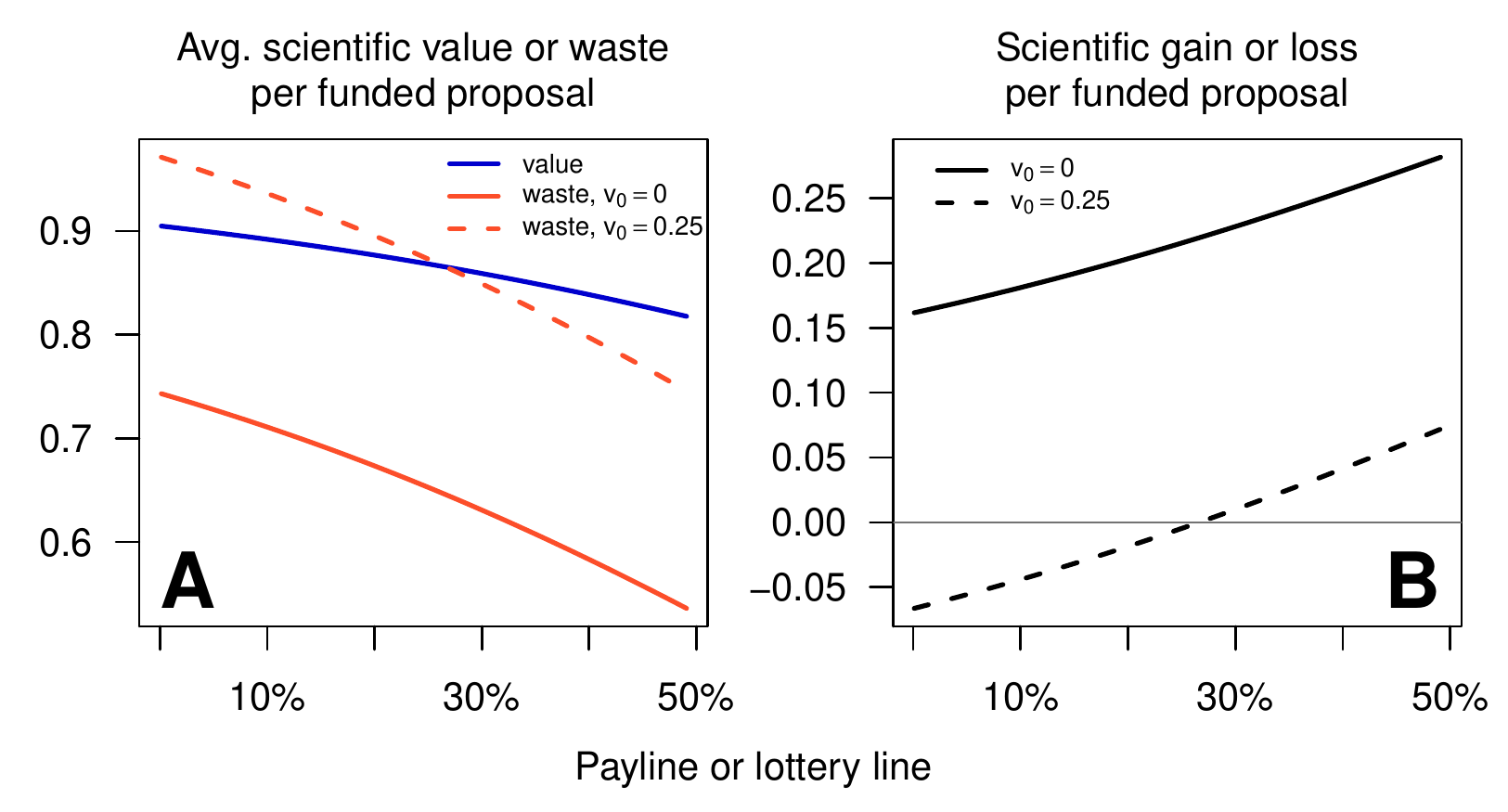}
		\caption{Parallel results to Fig.~\ref{fig:principal-a}, except with the first alternative parameter set given in the SI.  Note that the vertical axis in panel A does not extend to 0.}
		\label{fig:principal-b}
	\end{center}
\end{figure}

\begin{figure}
	\begin{center}
		\includegraphics[width=0.8\linewidth]{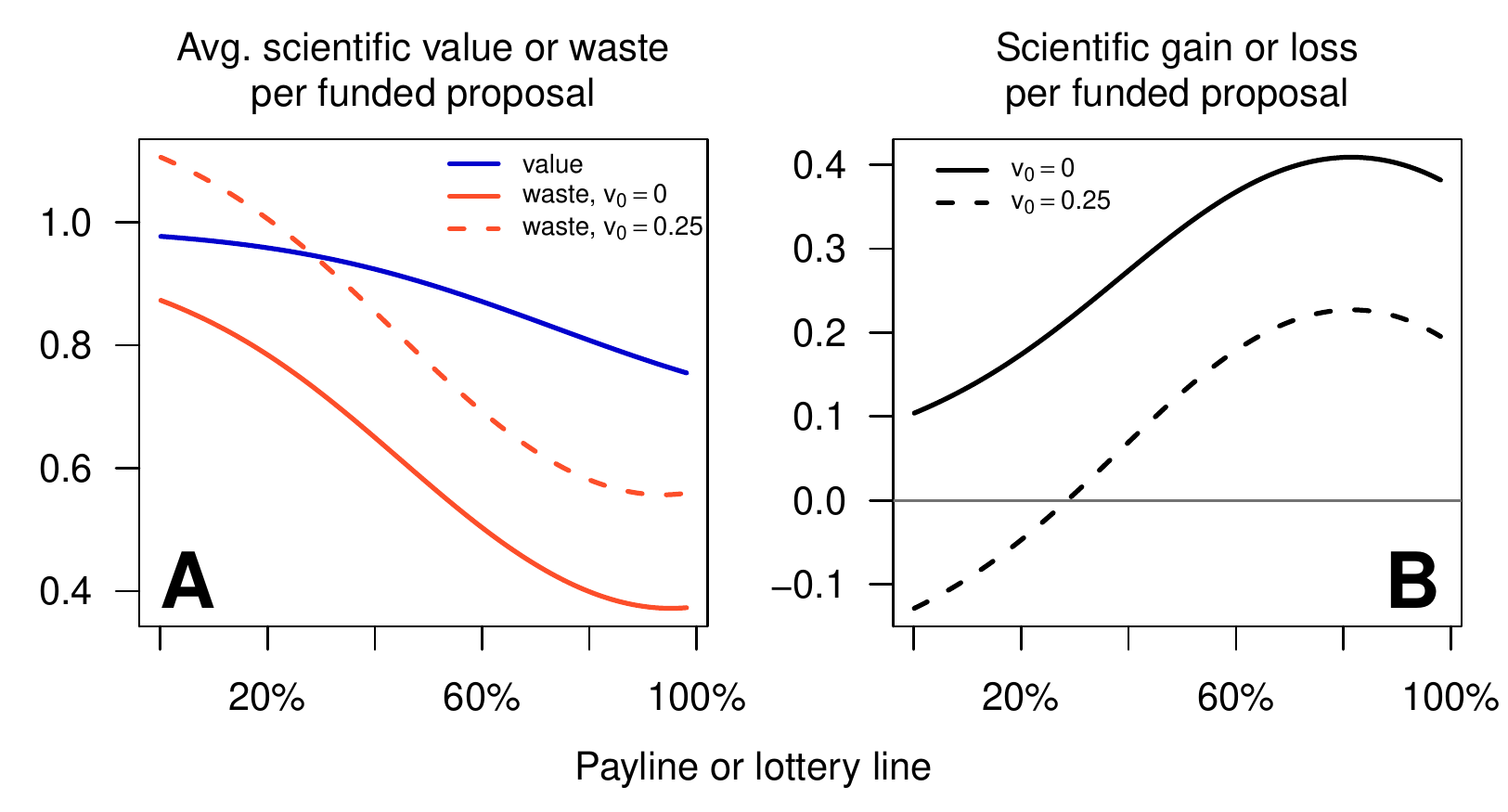}
		\caption{Parallel results to Fig.~\ref{fig:principal-a}, except with the second alternative parameter set given in the SI.  In this figure, data are shown for paylines ranging from $p=0.001$ to $p = 0.999$.  Note that the vertical axis in panel A does not extend to 0.}
		\label{fig:principal-c}
	\end{center}
\end{figure}

\begin{figure}
	\begin{center}
		\includegraphics[width=0.8\linewidth]{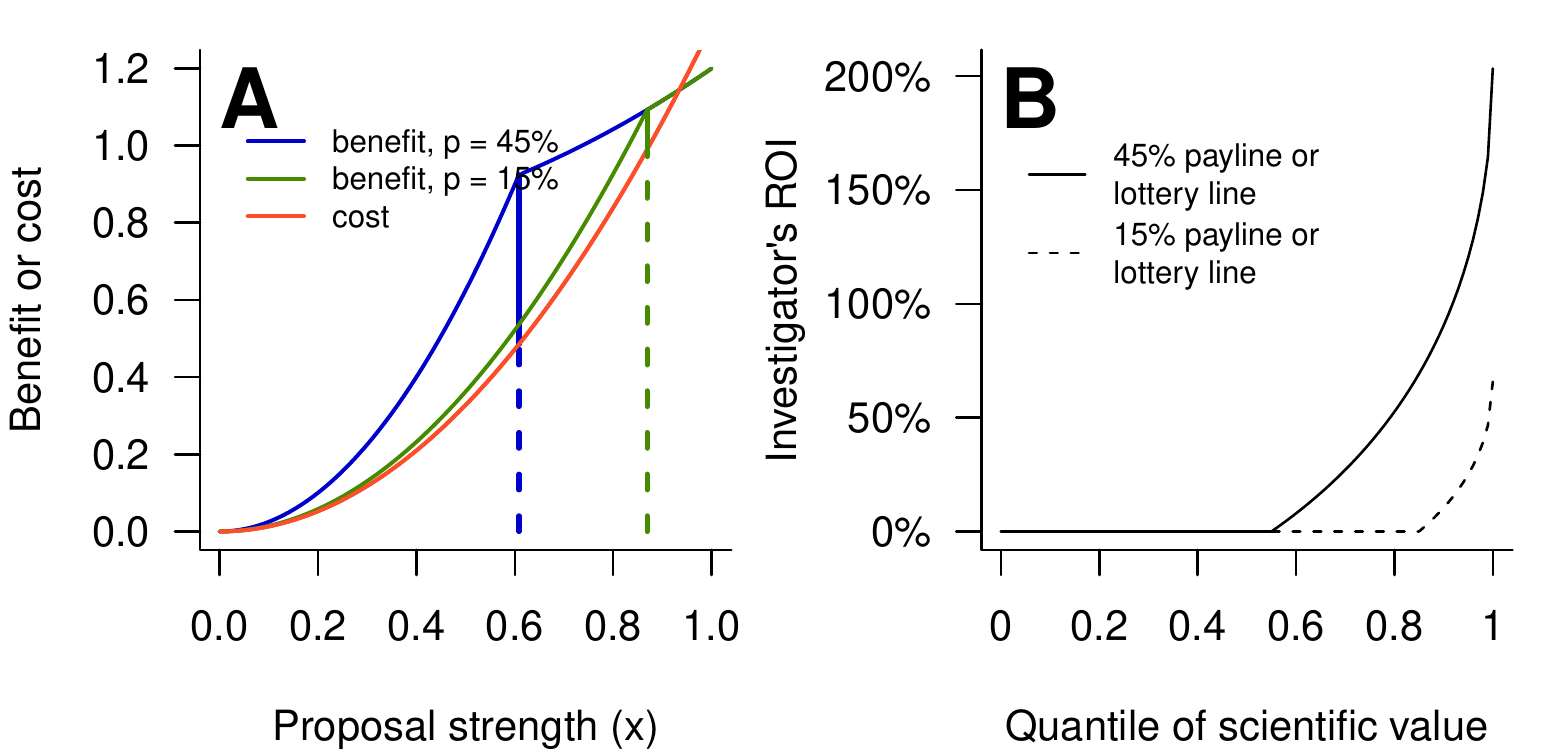}
		\caption{Parallel results to Fig.~\ref{fig:agent-a}, except with perfect assessment of grant quality.  All other parameter values are the same as in Fig.~\ref{fig:agent-a}.}
		\label{fig:agent-perfect}
	\end{center}
\end{figure}

\begin{figure}
	\begin{center}
		\includegraphics[width=0.8\linewidth]{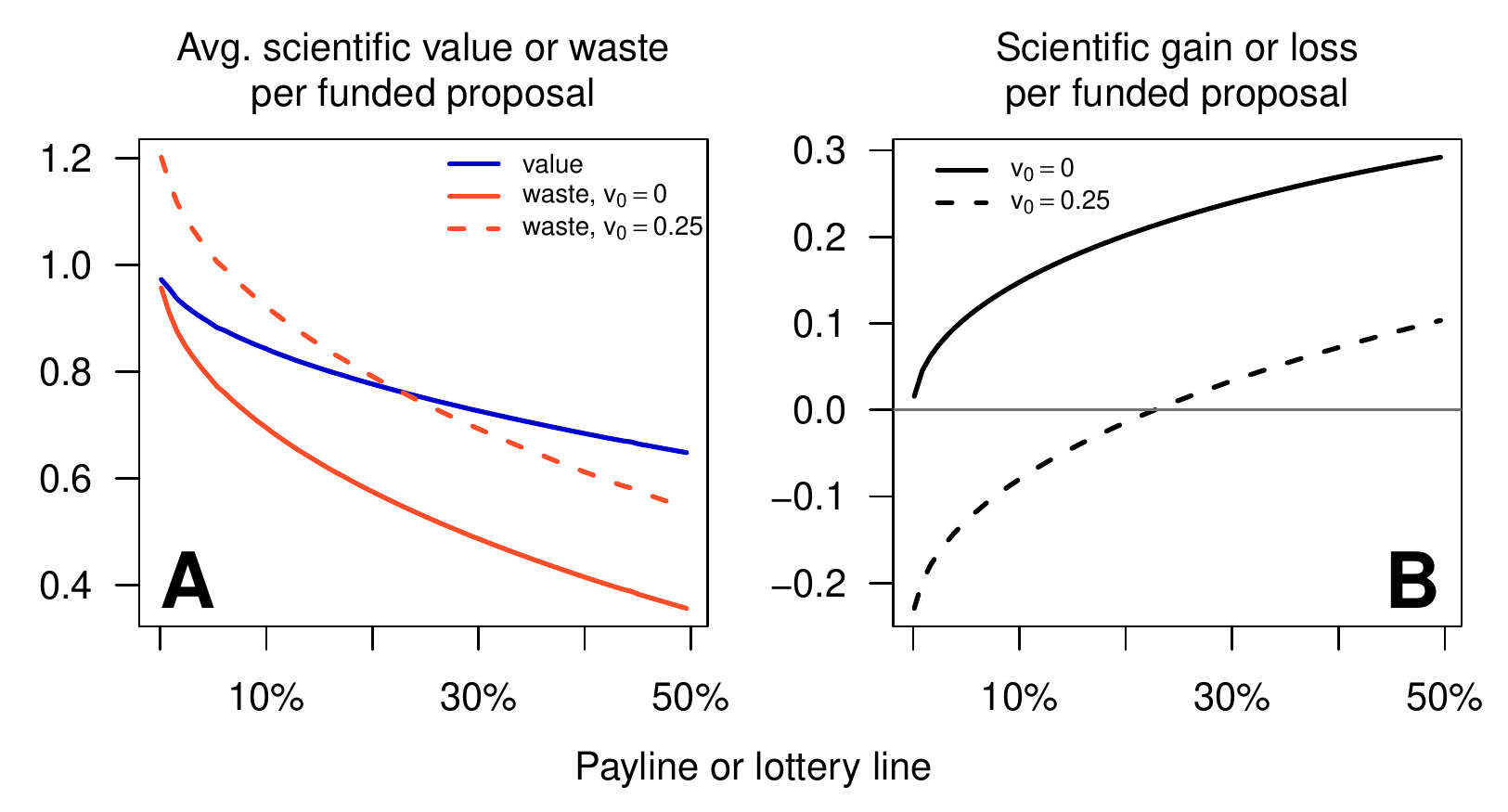}
		\caption{Parallel results to Fig.~\ref{fig:principal-a}, except with perfect assessment of proposal strength.  All other parameter values are the same as in Fig.~\ref{fig:principal-a}.  Note that the vertical axis in panel A does not extend to 0.}
		\label{fig:principal-perfect}
	\end{center}
\end{figure}

\begin{figure}
	\begin{center}
		\includegraphics[width=0.8\linewidth]{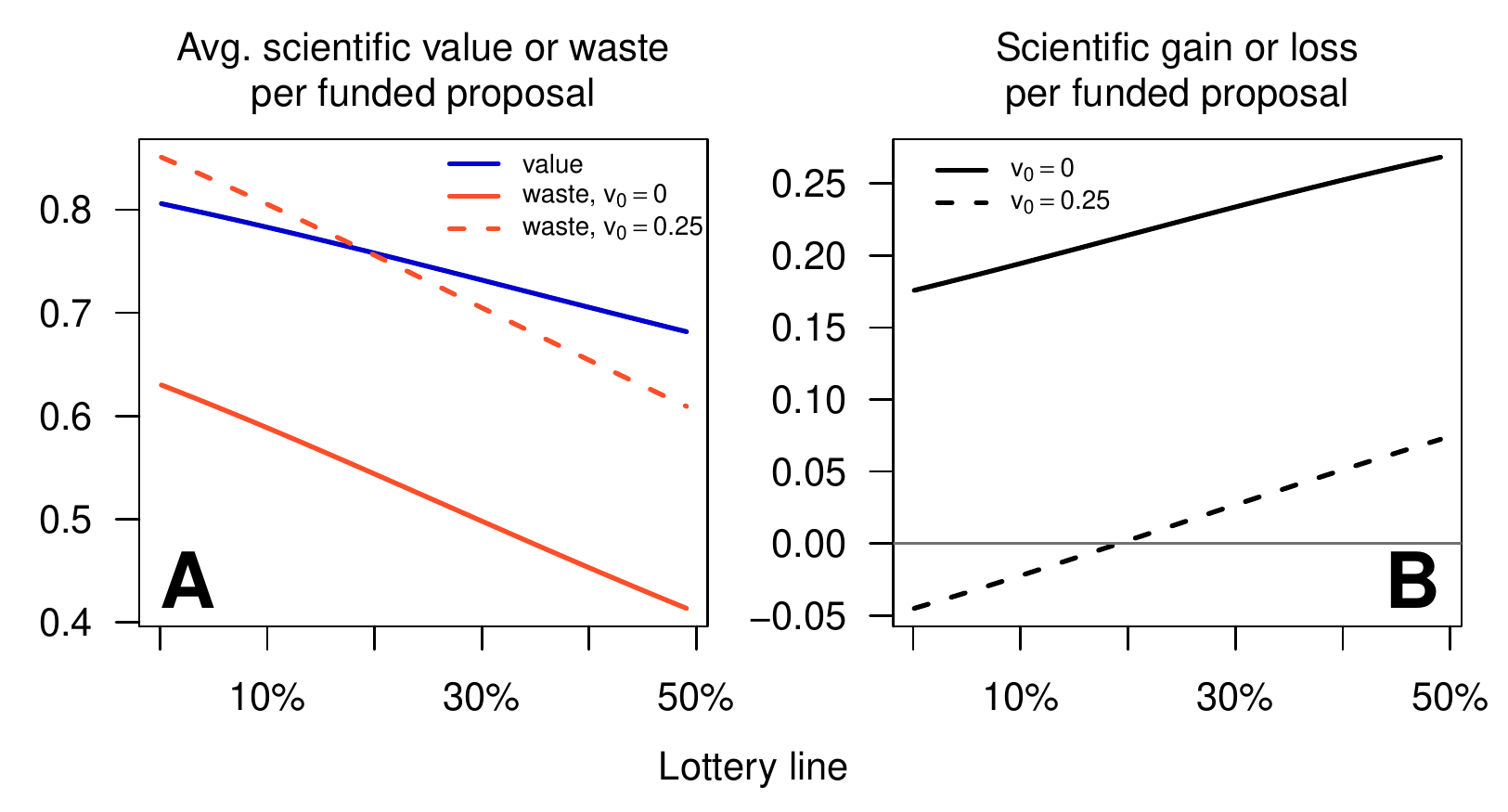}
		\caption{Parallel results to Fig.~\ref{fig:principal-a} for a three-tier lottery with equally sized tiers and a 3:2:1 ratio of funding probabilities across the tiers.  The horizontal axis gives the proportion of proposals that qualify for any tier of the lottery.  Scientific value and scientific waste per funded proposal are independent of the actual payline, as long as the payline is less than 2/3 of the lottery line.  (If the payline exceeds 2/3 of the lottery line, then the ratios of funding across tiers will be something other than 3:2:1, and thus the average value and average cost of a funded proposal will change slightly.)  All other parameter values are the same as in Fig.~\ref{fig:principal-a}.  Note that the vertical axis in panel A does not extend to 0.}
		\label{fig:principal-lottery}
	\end{center}
\end{figure}
\clearpage
\end{document}